\documentclass{elsart}

\usepackage{amsmath,amssymb}

\numberwithin{equation}{section}

\journal{Annals of Physics}
\newcommand{\cH}{\mathcal{H}}

\newcommand{\cN}{\mathcal{N}}

\newcommand{\cS}{\mathcal{S}}

\newcommand{\cV}{\mathcal{V}}
\newcommand{\cW}{\mathcal{W}}
\newcommand{\bA}{\boldsymbol{A}}
\newcommand{\bB}{\boldsymbol{B}}
\newcommand{\bC}{\boldsymbol{C}}

\newcommand{\bS}{\mathbf{S}}
\newcommand{\tH}{\tilde{\!H}{}}

\newcommand{\tcH}{\tilde{\cH}}
\newcommand{\tcV}{\tilde{\cV}}

\newcommand{\bi}{\bar{\imath}}
\newcommand{\bj}{\bar{\jmath}}
\newcommand{\bk}{\bar{k}}
\newcommand{\bl}{\bar{l}}

\newcommand{\bm}{\overline{m}}
\newcommand{\hz}{\hat{z}}

\newcommand{\bbZ}{\mathbb{Z}}
\newcommand{\bbR}{\mathbb{R}}
\newcommand{\bbC}{\mathbb{C}}
\newcommand{\ii}{\mathrm{i}}
\newcommand{\ee}{\mathrm{e}}
\newcommand{\dd}{\mathrm{d}}
\newcommand{\del}{\partial}

\newcommand{\rnu}{\sqrt{\nu}}

\newcommand{\fsl}{\mathfrak{sl}} 
\newcommand{\fgl}{\mathfrak{gl}} 
\newcommand{\gd}{\operatorname{gd}}

\begin{document}

\begin{frontmatter}


%
%

\title{A New Construction of Quasi-solvable Quantum Many-body Systems of
Deformed Calogero--Sutherland Type}
\author{Toshiaki Tanaka}
\ead{ttanaka@mail.tku.edu.tw}
\address{Departamento de F\'{\i}sica Te\'orica II,
  Facultad de Ciencias F\'{\i}sicas,\\
  Universidad Complutense, 28040 Madrid, Spain\thanksref{TT}}
\thanks[TT]{Present address: Department of Physics, Tamkang
University, Tamsui 25137, Taiwan, R.O.C.}


\begin{abstract}

We make a new multivariate generalization of the type A monomial space
of a single variable. It is different from the previously introduced
type A space of several variables which is an $\fsl(M+1)$ module, and we
thus call it type A$'$. We construct the most general quasi-solvable
operator of (at most) second-order which preserves the type A$'$ space.
Investigating directly the condition under which the type A$'$ operators
can be transformed to Schr\"odinger operators, we obtain the complete list
of the type A$'$ quasi-solvable quantum many-body systems. In particular,
we find new quasi-solvable models of deformed Calogero--Sutherland type
which are different from the Inozemtsev systems. We also examine a new
multivariate generalization of the type C monomial space based on
the type A$'$ scheme.

\end{abstract}




\end{frontmatter}

\section{Introduction}
\label{sec:intro}

\noindent
It is widely known that most of the quasi-exactly solvable quantum
one-body Hamiltonians, for which we can obtain a part of the exact
eigenvalues and eigenfunctions in closed form \cite{TU87,Us94},
have the underlying $\fsl(2)$ structure first discovered in
Ref.~\cite{Tu88}. Recently, it was found in Ref.~\cite{RT95} that
the well-known exactly solvable $M$-body Calogero--Sutherland models
\cite{Ca71,Su71} have a similar Lie-algebraic structure of $\fsl(M+1)$.
After the several new discoveries of quasi-exactly solvable
Hamiltonians having the same $\fsl(M+1)$ Lie-algebraic structure
\cite{MRT96,BTW98,HS99,GGR00}, the complete list of the quantum
many-body systems which admit the $\fsl(M+1)$ structure was obtained
in Ref.~\cite{Ta04}. It turns out that all of them are of Inozemtsev type,
which were originally found to be classically integrable
\cite{In83,IM85,In89}. The common feature of all the models which have
an underlying Lie-algebraic structure is that they leave a finite
dimensional module of the Lie algebra invariant.

On the other hand, Post and Turbiner studied a classification 
of linear differential operators of a single variable which have a finite
dimensional invariant subspace spanned by monomials \cite{PT95}. According
to the results in Ref.~\cite{PT95} combined with the discussion in
Ref.~\cite{GT04}, there are essentially three different spaces of
monomial type preserved by second-order linear differential operators,
except for a few special cases.
Later, they were dubbed type A, B, and C, respectively, according to
the corresponding types of $\cN$-fold supersymmetry \cite{GT05}.
The type A space corresponds to the $\fsl(2)$ module investigated in
Ref.~\cite{Tu88} and second-order linear differential operators preserving
it are expressed as quadratic forms of the $\fsl(2)$ generators
represented by first-order linear differential operators. The $\fsl(M+1)$
module preserved by the quantum Inozemtsev systems was then regarded as
a natural generalization of the type A monomial space of a single variable
to several variables \cite{Ta05a}.

However, the other two spaces, type B and C, are not Lie-algebraic modules
and linear differential operators preserving the type B or C spaces are
not given through the universal enveloping algebra of any Lie algebras.
Recently, we have successfully generalized the type C monomial space of
a single variable to several variables and constructed the most general
second-order linear differential operator preserving it \cite{Ta05a}.
Throughout our experiences in studying the latter problems and searching
for a natural generalization of the type B monomial space of a single
variable to several variables (cf. Section 8 in Ref.~\cite{Ta05a}), we are
convinced that there are much more varieties
of monomial type space of several variables than those of a single variable
which can be preserved by linear differential operators if we do not
restrict ourselves to study such an operator that admits an underlying
Lie-algebraic structure.

In this article, we show that by making another generalization of the type
A monomial space of a single variable, which is different from the
$\fsl(M+1)$ module investigated fully in Ref.~\cite{Ta04}, we obtain
a new family of quasi-solvable quantum many-body systems of deformed
Calogero--Sutherland type. The obtained models turn to have in general
$M$-body interaction terms and are thus different from the Inozemtsev
systems.

The article is organized as follows. In the next section, we summarize
some definitions related to the concept of quasi-solvability in order
to avoid ambiguity. In Section~\ref{sec:gentA}, we introduce a new
generalization of the type A monomial space of a single variable to
several variables, which we shall call type A$'$, and briefly discuss some
important property of the type A$'$ space such as the invariance
under linear transformations. In Section~\ref{sec:const},
we construct the most general (at most) second-order linear differential
operator which preserves the type A$'$ space. In Section~\ref{sec:extra},
we examine the condition under which the general type A$'$ quasi-solvable
operators can be transformed to Schr\"odinger operators.
Utilizing the invariance under the linear transformations, we fully
classify the possible quantum Hamiltonians preserving the type A$'$ space in
Section~\ref{sec:class}.
In Section~\ref{sec:newtC}, we make a new multivariate generalization of
the type C monomial space of a single variable, which we shall call type
C$'$, based on the type A$'$ space. We then construct the most general
type C$'$ quasi-solvable operator of (at most) second-order as well as
the type C$'$ gauged Hamiltonian.
Finally in Section~\ref{sec:discus}, we summarize and discuss the obtained
results in combination with the results in the type A and C cases. Some
useful formulas are summarized in Appendix~\ref{sec:formu}.

\section{Definition}
\label{sec:defin}

First of all, we shall give the definition of quasi-solvability and
some notions of its special cases based on Refs.~\cite{Ta04,Tu94}.
A linear differential operator $H$ of several variables
$q=(q_{1},\dots,q_{M})$ is said to be \textit{quasi-solvable}
if it preserves a finite dimensional functional space $\cV_{\cN}$
whose basis admits an analytic expression $\phi_{i}(q)$ in closed
form\footnote{The latter restriction has been sometimes missed in
the literature. Without it, however, arbitrary linear operators would
be quasi-solvable unless their spectrum is empty.}:
\begin{align}
\label{eq:defqs}
H\cV_{\cN}&\subset\cV_{\cN}, & \dim\cV_{\cN}&=n(\cN)<\infty,
 & \cV_{\cN}&=\bigl\langle\phi_{1}(q),\dots,\phi_{n(\cN)}(q)
 \bigr\rangle.
\end{align}
An immediate consequence of the above definition of quasi-solvability
is that, since we can calculate finite dimensional matrix elements
$\bS_{k,l}$ defined by,
\begin{align}
\label{eq:defbS}
H\phi_{k}=\sum_{l=1}^{n(\cN)}\bS_{k,l}\phi_{l}\qquad
 \bigl(k=1,\dots,n(\cN)\bigr),
\end{align}
we can diagonalize the operator $H$ and obtain its spectra
in the space $\cV_{\cN}$, at least, algebraically. Furthermore,
if the space $\cV_{\cN}$ is a subspace of a Hilbert space
$L^{2}(S)$ $(S\subset\bbR^{M})$ on which the operator $H$ is naturally
defined, the calculable spectra and the corresponding vectors in
$\cV_{\cN}$ give the \emph{exact} eigenvalues and eigenfunctions of
$H$, respectively. In this case, the operator $H$ is said to be
\emph{quasi-exactly solvable} (on $S$)\footnote{A domain $S$ is not
necessarily a subspace of the real space $\bbR^{M}$ if the operator under
consideration is non-Hermitian. Indeed, a couple of quasi-solvable one-body
Hamiltonians which are not quasi-exactly solvable on any subspaces of the
real space $\bbR$ are shown to be quasi-exactly solvable on the subspaces
of the \emph{complex} space $\bbC$ incorporating with the
$\mathcal{PT}$-symmetric boundary conditions \cite{BB98,BM05}.}.
Otherwise, the calculable spectra and the corresponding vectors in
$\cV_{\cN}$ only give \emph{local} solutions of the characteristic
equation of $H$. This important difference has been sometimes missed
in the literature.

A quasi-solvable operator $H$ of several variables is said to be
\emph{solvable} if it preserves an infinite flag of finite dimensional
functional spaces $\cV_{\cN}$,
\begin{align}
\label{eq:flagv}
\cV_{1}\subset\cV_{2}\subset\dots\subset\cV_{\cN}\subset\cdots,
\end{align}
whose bases admit analytic expressions in closed form, that is,
\begin{align}
\label{eq:defsv}
H\cV_{\cN}&\subset\cV_{\cN}, & \dim\cV_{\cN}&=n(\cN)<\infty,
 & \cV_{\cN}&=\bigl\langle\phi_{1}(q),\dots,\phi_{n(\cN)}(q)
 \bigr\rangle,
\end{align}
for $\cN=1,2,3,\ldots$.
Furthermore, if the sequence of the spaces \eqref{eq:flagv} defined
on $S\subset\bbR^{M}$ satisfies,
\begin{align}
\label{eq:limvs}
\overline{\cV_{\cN}(S)}\rightarrow L^{2}(S)\qquad
 (\cN\rightarrow\infty),
\end{align}
the operator $H$ is said to be \emph{exactly solvable} (on $S$).

\section{A New Generalization of the Type A Monomial Space}
\label{sec:gentA}

In this article, we shall consider a quantum mechanical system of $M$
identical particles on a line. The Hamiltonian is thus given by
\begin{align}
\label{eq:mbHam}
H=-\frac{1}{2}\sum_{i=1}^{M}\frac{\del^{2}}{\del q_{i}^{2}}
 +V(q_{1},\dots,q_{M}),
\end{align}
where the potential has permutation symmetry:
\begin{align}
V(\dots,q_{i},\dots,q_{j},\dots)=V(\dots,q_{j},\dots,q_{i},\dots)
 \qquad\forall i\neq j\,.
\end{align}
To construct a quasi-solvable operator of the form \eqref{eq:mbHam},
we shall follow the three steps after Ref.~\cite{Ta04}, namely, i) a
\emph{gauge} transformation on the Hamiltonian \eqref{eq:mbHam}:
\begin{align}
\label{eq:gtHam}
\tH=\ee^{\cW(q)}H\ee^{-\cW(q)},
\end{align}
ii) a change of variables from $q_{i}$ to $z_{i}$ by a function $z$ of a single
variable:
\begin{align}
z_{i}=z(q_{i}),
\end{align}
and iii) the introduction of the elementary symmetric polynomials of $z_{i}$
defined by,
\begin{align}
\label{eq:espol}
\sigma_k(z)=\sum_{i_{1}<\dots<i_{k}}^{M}z_{i_{1}}\dots z_{i_{k}}
 \quad (k=1,\dots,M),\qquad\sigma_{0}\equiv 1\,.
\end{align}
Due to the permutation symmetry of the original Hamiltonian \eqref{eq:mbHam},
the gauged Hamiltonian \eqref{eq:gtHam} can be completely expressed in terms
of the elementary symmetric polynomials \eqref{eq:espol}.
In this article, a second-order linear differential operator is called
a \emph{gauged Hamiltonian} if it can be transformed to a Schr\"odinger
operator by means of a combination of gauge transformations and change
of variables.

The next task is to choose a vector space to be preserved by the gauged
Hamiltonian \eqref{eq:gtHam}. The type A monomial space of a single variable
$z$ is defined by~\cite{GT05}
\begin{align}
\label{eq:stypA}
\tcV_{\cN}^{(A)}=\left\langle 1,z,\dots,z^{\cN-1}\right\rangle.
\end{align}
It is an $\fsl(2)$ module and the foundation of the $\fsl(2)$ construction
of quasi-solvable models in Ref.~\cite{Tu88}. In our previous
paper~\cite{Ta05a}, we identified the following space as a generalization
of the type A space of a single variable to several variables:
\begin{align}
\label{eq:mtypA}
\tcV_{\cN;M}^{(A)}=\Bigl\langle \sigma_{1}^{n_{1}}\dots\sigma_{M}^{n_{M}}
 \Bigm| n_{i}\in\bbZ_{\geq 0},\ \sum_{i=1}^{M}n_{i}\leq\cN-1\Bigr\rangle.
\end{align}
It is indeed a natural generalization since it provides an $\fsl(M+1)$
module for each $M$ and is the foundation of the $\fsl(M+1)$ construction
of quasi-solvable quantum many-body systems in Ref.~\cite{Ta04}.
However, the spaces \eqref{eq:stypA} and \eqref{eq:mtypA} have a
different character, that is, the elements of the latter space of
multi-variable are not necessarily polynomials of degree less than $\cN$
in the variables $z_{i}$, in contrast to the former space of a single
variable. Hence, another natural generalization would be such that
any element of a generalized space is a polynomial of degree less than
$\cN$ in $z_{i}$. Each elementary symmetric polynomial $\sigma_{k}$
is a polynomial of degree $k$ in $z_{i}$ and thus the latter
generalization can be realized by the following vector space:
\begin{align}
\label{eq:mtypA'}
\tcV_{\cN;M}^{(A')}=\Bigl\langle \sigma_{1}^{n_{1}}\dots\sigma_{M}^{n_{M}}
 \Bigm| n_{i}\in\bbZ_{\geq 0},\ \sum_{i=1}^{M}i n_{i}\leq\cN-1\Bigr\rangle.
\end{align}
Obviously, the space \eqref{eq:mtypA'} also reduces to the single-variable
type A space \eqref{eq:stypA} when $M=1$, but is different from
Eq.~\eqref{eq:mtypA}. We thus call the space \eqref{eq:mtypA'}
\emph{type A$'$}. In this article, we investigate linear differential
operators which preserve the type A$'$ space.

In contrast to the fact that the type A space \eqref{eq:mtypA} is invariant
under the $GL(2,\bbR)$ linear fractional transformation \cite{Ta04}:
\begin{align}
\tcV_{\cN}^{(A)}[z]\mapsto\prod_{i=1}^{M}(\gamma z_{i}+\delta)^{\cN-1}
 \tcV_{\cN}^{(A)}[\hz]=\tcV_{\cN}^{(A)}[z],
\end{align}
induced by
\begin{align}
\label{eq:fract}
z_{i}\mapsto\hz_{i}=\frac{\alpha z_{i}+\beta}{\gamma z_{i}+\delta}\qquad
 (\alpha,\beta,\gamma,\delta\in\bbR;\Delta\equiv\alpha\delta-\beta\gamma
 \neq 0),
\end{align}
the type A$'$ space \eqref{eq:mtypA'} does not have the full $GL(2,\bbR)$
invariance for $M>1$. It is easily read from the transformation of the
elementary symmetric polynomials \eqref{eq:espol} under the special
projective transformation $z_{i}\mapsto\hz_{i}=1/z_{i}$:
\begin{align}
\sigma_{k}(z)\mapsto\sigma_{k}(z^{-1})=\sigma_{M-k}(z)\sigma_{M}(z)^{-1}.
\end{align}
Hence, the special projective transformation interchanges the role of
the variables $\sigma_{k}$ and $\sigma_{M-k}$ in the space \eqref{eq:mtypA'}
and cannot keep the condition $\sum_{i=1}^{M}i n_{i}\leq\cN-1$ unchanged.
However, the other elements of the $GL(2,\bbR)$ transformations
\eqref{eq:fract}, namely, the dilatation ($\hz_{i}=\alpha z_{i}$) and
the translation ($\hz_{i}=z_{i}+\beta$), preserve the type A$'$ space
\eqref{eq:mtypA'}. The former is trivial while the latter is understood
from the transformation of $\sigma_{k}$ under the translation:
\begin{align}
\sigma_{k}(z)\mapsto\sigma_{k}(z+\beta)=\sum_{l=0}^{k}\beta^{k-l}
 C_{l}(M)\sigma_{l}(z),
\end{align}
where $C_{l}$ are constants depending on $M$. We now easily see that
the latter transformation indeed preserves all the elements of
\eqref{eq:mtypA'} within the space.

\section{Construction of Quasi-solvable Operators}
\label{sec:const}

In this section, we shall construct the general quasi-solvable operators
of (at most) second-order leaving the type A$'$ space \eqref{eq:mtypA'}
invariant. The set of linearly independent first-order differential
operators preserving the type A$'$ space is given by,
\begin{subequations}
\label{eqs:1st}
\begin{align}
\label{eq:1st1}
F_{\{m_{i}\}_{k},k}
 &\equiv\prod_{i=1}^{M}\sigma_{i}^{m_{i}}\frac{\del}{\del\sigma_{k}}
 \quad(k=1,\dots,M),\\
\label{eq:1st2}
F_{10}&\equiv\sigma_{1}\biggl(\cN-1-\sum_{k=1}^{M}k\sigma_{k}
 \frac{\del}{\del\sigma_{k}}\biggr).
\end{align}
\end{subequations}
In Eq.~\eqref{eq:1st1}, $\{m_{i}\}_{k}$ is an abbreviation of the set of
$M$ non-negative integers defined by
\begin{align}
\label{eq:set1}
\{m_{i}\}_{k}\equiv\Bigl\{m_{1},\dots,m_{M}\Bigm| m_{i}\in\bbZ_{\geq 0},
 \ \sum_{i=1}^{M}i m_{i}\leq k\Bigr\}.
\end{align}
In the single variable case ($M=1$), the set of differential operators
\eqref{eqs:1st} consists of
\begin{align}
F_{\{0\}_{1},1}=\frac{\del}{\del\sigma_{1}},\quad
 F_{\{1\}_{1},1}=\sigma_{1}\frac{\del}{\del\sigma_{1}},\quad
 F_{10}=\sigma_{1}\biggl(\cN-1-\sigma_{1}\frac{\del}{\del\sigma_{1}}\biggr),
\end{align}
and hence is essentially the same as the $\fsl(2)$ generators. It is
as expected since the single-variable type A$'$ space is nothing but
the $\fsl(2)$ module \eqref{eq:stypA}. In the case of two variables ($M=2$),
the set of differential operators \eqref{eqs:1st} is composed of
\begin{align}
F_{\{00\}_{1},1}&=\frac{\del}{\del\sigma_{1}},\quad
 F_{\{10\}_{1},1}=\sigma_{1}\frac{\del}{\del\sigma_{1}},\quad
 F_{10}=\sigma_{1}\biggl(\cN-1-\sum_{k=1}^{2}k\sigma_{k}
 \frac{\del}{\del\sigma_{k}}\biggr),\notag\\
\label{eq:1stM2}
F_{\{01\}_{2},2}&=\sigma_{2}\frac{\del}{\del\sigma_{2}},\quad
 F_{\{m_{1}0\}_{2},2}=\sigma_{1}^{m_{1}}\frac{\del}{\del\sigma_{2}}\quad
 (m_{1}=0,1,2),
\end{align}
and is essentially the same as the generators of $\fgl(2)\ltimes\bbR^{3}$ in
Ref.~\cite{Tu94}. Actually, the two-variable type A$'$ space \eqref{eq:mtypA'}
coincides with the $\fgl(2)\ltimes\bbR^{3}$ module investigated in
Ref.~\cite{Tu94}. In this sense, the type A$'$ provides a natural extension of
that Lie-algebraic scheme of two variables to arbitrary number of variables.

The set of linearly independent second-order differential operators leaving
the type A$'$ space invariant is as follows:
\begin{align}
\label{eq:2nd1}
F_{\{m_{i}\}_{k+l},kl}&\equiv\prod_{i=1}^{M}\sigma_{i}^{m_{i}}
 \frac{\del^{2}}{\del\sigma_{k}\del\sigma_{l}}\quad (k,l=1,\dots,M;\,
 k\geq l),\\
\label{eq:2nd2}
F_{10}F_{\{\bm_{i}\}_{k},k}&=\sigma_{1}\biggl(\cN-1-\sum_{l=1}^{M}l
 \sigma_{l}\frac{\del}{\del\sigma_{l}}\biggr)\prod_{i=1}^{M}
 \sigma_{i}^{\bm_{i}}\frac{\del}{\del\sigma_{k}}\quad(k=1,\dots,M),\\
F_{10}F_{10}&=\sigma_{1}^{2}\biggl(\cN-2-\sum_{k=1}^{M}k\sigma_{k}
 \frac{\del}{\del\sigma_{k}}\biggr)\biggl(\cN-1-\sum_{l=1}^{M}l\sigma_{l}
 \frac{\del}{\del\sigma_{l}}\biggr),\\
\label{eq:2nd4}
F_{20,00}&\equiv\sigma_{2}\biggl(\cN-2-\sum_{k=1}^{M}k\sigma_{k}
 \frac{\del}{\del\sigma_{k}}\biggr)\biggl(\cN-1-\sum_{l=1}^{M}l\sigma_{l}
 \frac{\del}{\del\sigma_{l}}\biggr),
\end{align}
where in Eq.~\eqref{eq:2nd2} $\{\bm_{i}\}_{k}$ is an abbreviation of
the set of $M$ non-negative integers defined by
\begin{align}
\label{eq:set2}
\{\bm_{i}\}_{k}\equiv\Bigl\{\bm_{1},\dots,\bm_{M}\Bigm|
 \bm_{i}\in\bbZ_{\geq 0},\ \sum_{i=1}^{M}i \bm_{i}=k\Bigr\}.
\end{align}
Several remarks are in order. First of all, it is important to note that
the set of the quadratic form of the first-order operators
$F_{\{m_{i}\}_{k},k}$ in Eq.~\eqref{eq:1st1} cannot exhaust
the second-order operators of the form given by Eq.~\eqref{eq:2nd1}
when the number of the variables is greater than two ($M>2$).
In the case of $M=3$, for instance, the following operator which is
an element of Eq.~\eqref{eq:2nd1}
\begin{align}
F_{\{020\}_{4},31}
 =\sigma_{2}^{2}\frac{\del^{2}}{\del\sigma_{3}\del\sigma_{1}},
\end{align}
cannot be represented by any quadratic combination of the first-order
operators in Eq.~\eqref{eq:1st1}. Second, even when the number of
the variables is two ($M=2$) where the type A$'$ space \eqref{eq:mtypA'}
provides a $\fgl(2)\ltimes\bbR^{3}$ module, there exist higher-order
operators preserving the type A$'$ space which cannot be expressed as
a polynomial in the first-order operators \eqref{eq:1stM2}, such as
Eq.~\eqref{eq:2nd4}. In this respect, see also Ref.~\cite{CRT98}.
Third, the operators $F_{10}F_{\{m_{i}\}_{k},k}$ with
$\sum_{i=1}^{M}im_{i}<k$ are represented by linear combinations of
the other operators \eqref{eqs:1st}--\eqref{eq:2nd4} as
\begin{align}
F_{10}F_{\{m_{i}\}_{k},k}
&=\biggl(\cN-1-\sum_{l=1}^{M}lm_{l}\biggr)\sigma_{1}\prod_{i=1}^{M}
 \sigma_{i}^{m_{i}}\frac{\del}{\del\sigma_{k}}-\sum_{l=1}^{M}l\sigma_{1}
 \sigma_{l}\prod_{i=1}^{M}\sigma_{i}^{m_{i}}
 \frac{\del^{2}}{\del\sigma_{k}\del\sigma_{l}}\notag\\
&=\biggl(\cN-1-\sum_{l=1}^{M}lm_{l}\biggr)F_{\{m'_{i}\}_{k},k}
 -\sum_{l=1}^{M}l F_{\{m''_{i}\}_{k+l},kl}\,,
\end{align}
where $\{m'_{i}\}_{k}$ is obtained from $\{m_{i}\}_{k}$ by
$m'_{i}=m_{i}+\delta_{i,1}$ while $\{m''_{i}\}_{k+l}$ is from
$\{m_{i}\}_{k}$ by $m''_{i}=m_{i}+\delta_{i,1}+\delta_{i,l}$.
The restriction $\sum_{i=1}^{M}im_{i}<k$ ensures that
$\sum_{i=1}^{M}im'_{i}\leq k$ and $\sum_{i=1}^{M}im''_{i}\leq k+l$.
Thus, $F_{\{m'_{i}\}_{k},k}$ and $F_{\{m''_{i}\}_{k+l},kl}$ are in fact
members of the operators in Eqs.~\eqref{eq:1st1} and \eqref{eq:2nd1},
respectively. Hence, the operators $F_{10}F_{\{m_{i}\}_{k},k}$ with
$\sum_{i=1}^{M}im_{i}<k$ are linearly dependent on the other operators
\eqref{eqs:1st}--\eqref{eq:2nd4}. That is why in Eq.~\eqref{eq:2nd2}
we restrict the values of the set $\{m_{i}\}$ to the one given by
Eq.~\eqref{eq:set2}.
Furthermore, the operators of the form $F_{\{\bm_{i}\}_{k},k}F_{10}$
are also linearly dependent on the set of operators
\eqref{eqs:1st}--\eqref{eq:2nd4} since the anti-commutator of
$F_{\{\bm_{i}\}_{k},k}$ and $F_{10}$ reads
\begin{align}
\bigl[ F_{\{\bm_{i}\}_{k},k}, F_{10}\bigr]=\delta_{k,1}F_{10}.
\end{align}
Therefore, the most general quasi-solvable operator of (at most)
second-order which preserves the type A$'$ space \eqref{eq:mtypA'} is
given by the linear combination of the operators
\eqref{eqs:1st}--\eqref{eq:2nd4}:
\begin{align}
\tcH_{\cN}^{(A')}=&\,-\sum_{k\geq l}^{M}\sum_{\{m_{i}\}_{k+l}}
 A_{\{m_{i}\}_{k+l},kl}F_{\{m_{i}\}_{k+l},kl}-\sum_{k=1}^{M}
 \sum_{\{\bm_{i}\}_{k}}A_{10,\{\bm_{i}\}_{k},k}F_{10}
 F_{\{\bm_{i}\}_{k},k}\notag\\
&\,-A_{10,10}F_{10}F_{10}-A_{20,00}F_{20,00}\notag\\
&\,+\sum_{k=1}^{M}\sum_{\{m_{i}\}_{k}}B_{\{m_{i}\}_{k},k}
 F_{\{m_{i}\}_{k},k}+B_{10}F_{10}-c_{0},
\end{align}
where the coefficients $A$, $B$ with indices and $c_{0}$ are real constants
and the summation over the set $\{m_{i}\}_{k}$ etc. is understood to take
all the possible set of values $\{m_{1},\dots,m_{M}\}$ indicated in
Eqs.~\eqref{eq:set1} and \eqref{eq:set2}. In terms of the variables
$\sigma$, the operator $\tcH_{\cN}^{(A')}$ is expressed as
\begin{align}
\label{eq:tA'op2}
\tcH_{\cN}^{(A')}=&\,-\sum_{k,l=1}^{M}\bigl[\bA_{0}(\sigma)kl\sigma_{k}
 \sigma_{l}+\bA_{kl}(\sigma)\bigr]
 \frac{\del^{2}}{\del\sigma_{k}\del\sigma_{l}}\notag\\
&\,+\sum_{k=1}^{M}\bigl[\bB_{0}(\sigma)k\sigma_{k}-\bB_{k}(\sigma)\bigr]
 \frac{\del}{\del\sigma_{k}}-\bC(\sigma),
\end{align}
where $\bA_{0}$, $\bA_{kl}$, $\bB_{0}$, $\bB_{k}$, and $\bC$ are polynomials
of several variables given by
\begin{subequations}
\begin{align}
\label{eq:polA0}
\bA_{0}(\sigma)=&\,A_{10,10}\sigma_{1}^{2}+A_{20,00}\sigma_{2},\\
\label{eq:polAkl}
\bA_{kl}(\sigma)=&\,\sum_{\{m_{i}\}_{k+l}}A_{\{m_{i}\}_{k+l},kl}
 \prod_{i=1}^{M}\sigma_{i}^{m_{i}}-\sum_{\{\bm_{i}\}_{k}}
 A_{10,\{\bm_{i}\}_{k},k}l\sigma_{1}\sigma_{l}\prod_{i=1}^{M}
 \sigma_{i}^{\bm_{i}},\\
\bB_{0}(\sigma)=&\,(2\cN-k-3)\bA_{0}(\sigma)-B_{10}\sigma_{1},\\
\bB_{k}(\sigma)=&\,(\cN-k-1)\sum_{\{\bm_{i}\}_{k}}A_{10,\{\bm_{i}\}_{k},k}
 \sigma_{1}\prod_{i=1}^{M}\sigma_{i}^{\bm_{i}}
 -\sum_{\{m_{i}\}_{k}}B_{\{m_{i}\}_{k},k}\prod_{i=1}^{M}\sigma_{i}^{m_{i}},\\
\label{eq:polC}
\bC(\sigma)=&\,(\cN-1)(\cN-2)\bA_{0}(\sigma)-(\cN-1)B_{10}\sigma_{1}+c_{0}.
\end{align}
\end{subequations}
Among the set of the operators \eqref{eqs:1st}--\eqref{eq:2nd4},
$F_{\{m_{i}\}_{k},k}$ and $F_{\{m_{i}\}_{k+l},kl}$ preserve the type A$'$
space \eqref{eq:mtypA'} for \emph{arbitrary} natural number $\cN$.
Therefore, the operator $\tcH_{\cN}^{(A')}$ is not only quasi-solvable but
also solvable if
\begin{align}
\label{eq:solv1}
A_{10,\{\bm_{i}\}_{k},k}=A_{10,10}=A_{20,00}=B_{10}=0.
\end{align}

\section{Extraction of Schr\"odinger Operators}
\label{sec:extra}

In the preceding section, we have constructed the most general
quasi-solvable second-order operator $\tcH_{\cN}^{(A')}$ preserving
the type A$'$ space \eqref{eq:mtypA'}. By applying a similarity
transformation on $\tcH_{\cN}^{(A')}$ and a change of variables,
we may obtain a family of quasi-solvable operators of a desired form.
However, second-order linear differential operators of several variables
are in general not gauged Hamiltonians, that is, they cannot be always
transformed to Schr\"odinger operators. This fact is one of the most
obstacles in constructing quasi-solvable \emph{quantum} many-body systems.
Recently in Refs.~\cite{Ta04,Ta05a,Ta03c}, it was shown that the amount of
the difficulty can be significantly reduced by considering the underlying
symmetry of the invariant space of quasi-solvable operators. In the present
case, however, there is no full $GL(2,\bbR)$ invariance, especially no
invariance under the special projective transformation, as we have mentioned
previously in Section~\ref{sec:gentA}. It turns out that the dilatation
and translation invariance are insufficient to extract the general form of
the type A$'$ gauged Hamiltonians, namely, the operators which can be
transformed to Schr\"odinger operators from the most general type A$'$
quasi-solvable second-order operators \eqref{eq:tA'op2}.

Let us first review the general condition under which a second-order
linear differential operator of several variables can be cast into
a Schr\"odinger operator \cite{Ta03c}. Suppose the operator under
consideration $\tcH$ has the following form in variables $z$:
\begin{align}
\label{eq:g2op}
\tcH=-\sum_{i,j=1}^{M}P_{ij}(z)\frac{\del^{2}}{\del z_{i}\del z_{j}}
 +\sum_{i=1}^{M}S_{i}(z)\frac{\del}{\del z_{i}}-T(z).
\end{align}
Then, it is readily shown that $\tcH$ can be cast into a Schr\"odinger
operator by a gauge transformation and a change of variables
$z_{i}=z(q_{i})$ such that
\begin{align}
\ee^{-\cW}\tcH\ee^{\cW}=-\frac{1}{2}\sum_{i=1}^{M}
 \frac{\del^{2}}{\del q_{i}^{2}}+V(q),
\end{align}
if and only if the following conditions are satisfied:
\begin{gather}
\label{eq:cond1}
P_{ij}(z)+P_{ji}(z)=0\quad(i>j),\\[6pt]
\label{eq:cond2}
(z'_{i})^{2}=2P_{ii}(z),\\[6pt]
\label{eq:cond3}
\frac{\del\cW}{\del q_{i}}=\frac{S_{i}(z)}{z'_{i}}+\frac{z''_{i}}{2z'_{i}},
\end{gather}
where $z'_{i}$ denotes the derivative of $z_{i}$ with respect to $q_{i}$.
The first condition \eqref{eq:cond1} in general consists of a set of
algebraic identities. The second and third conditions
\eqref{eq:cond2}--\eqref{eq:cond3} on the other hand are sets of
differential equations and do not necessarily have a solution. In order
that the second condition has a solution, the r.h.s. of Eq.~\eqref{eq:cond2}
must depend only on the single variable $z_{i}$:
\begin{align}
\label{eq:cond4}
(z'_{i})^{2}=2P_{ii}(z)=2A(z_{i}),
\end{align}
since we have assumed that the change of variables is determined by
a single function of a single variable $z_{i}=z(q_{i})$ and thus
the l.h.s. of Eq.~\eqref{eq:cond2} depends solely on the variable $q_{i}$.
Furthermore, in order that the third condition \eqref{eq:cond3} has
a solution, the following integrability condition must be fulfilled
for all $i\neq l$:
\begin{align}
\label{eq:cond5}
\frac{\del}{\del q_{l}}\frac{\del\cW}{\del q_{i}}
 =\frac{\del}{\del q_{i}}\frac{\del\cW}{\del q_{l}}
 \quad\Leftrightarrow\quad
\frac{1}{A(z_{i})}\frac{\del S_{i}(z)}{\del z_{l}}
=\frac{1}{A(z_{l})}\frac{\del S_{l}(z)}{\del z_{i}},
\end{align}
where Eq.~\eqref{eq:cond4} is employed. Therefore, the operator $\tcH$
can be transformed to a Schr\"odinger operator if it has the form of
\begin{align}
\label{eq:gaugH}
\tcH=-\sum_{i=1}^{M}A(z_{i})\frac{\del^{2}}{\del z_{i}^{2}}
 +\sum_{i=1}^{M}S_{i}(z)\frac{\del}{\del z_{i}}-T(z),
\end{align}
with $A(z_{i})$ and $S_{i}(z)$ satisfying Eqs.~\eqref{eq:cond4} and
\eqref{eq:cond5}.

Next, we note that in our present case we have made the additional
change of variables from $z$ to $\sigma$, Eq.~\eqref{eq:espol}.
{}From the formula \eqref{eq:chain}, the part of the second-order operators
in Eq.~\eqref{eq:tA'op2} has the following form in terms of $z$:
\begin{align}
-\sum_{k,l=1}^{M}\bigl[\bA_{0}(\sigma)kl\sigma_{k}\sigma_{l}
 +\bA_{kl}(\sigma)\bigr]\frac{\del^{2}}{\del\sigma_{k}\del\sigma_{l}}
=-\sum_{i,j=1}^{M}P_{ij}(z)\frac{\del^{2}}{\del z_{i}\del z_{j}},
\end{align}
with
\begin{align}
\label{eq:Pij}
P_{ij}(z)=\frac{\sum_{k,l=1}^{M}(-1)^{k+l}z_{i}^{M-k}z_{j}^{M-l}
 \bigl[\bA_{0}(\sigma)kl\sigma_{k}\sigma_{l}+\bA_{kl}(\sigma)\bigr]}{
 \prod_{m(\neq i)}^{M}(z_{i}-z_{m})\prod_{n(\neq j)}^{M}(z_{j}-z_{n})}.
\end{align}
To satisfy the condition \eqref{eq:cond4}, it is evident that the denominator
in the r.h.s. of Eq.~\eqref{eq:Pij} for $P_{ii}(z)$ must be completely
canceled with a part of the numerator; otherwise, $P_{ii}(z)$ cannot be
a function of the single variable $z_{i}$. As a consequence, the part of
the second-order operators in Eq.~\eqref{eq:tA'op2} which satisfy
the conditions \eqref{eq:cond1} and \eqref{eq:cond4} must have the
following form:
\begin{align}
\label{eq:2ndop2}
-\sum_{k,l=1}^{M}\bigl[\bA_{0}(\sigma)kl\sigma_{k}\sigma_{l}
 +\bA_{kl}(\sigma)\bigr]\frac{\del^{2}}{\del\sigma_{k}\del\sigma_{l}}
=-\sum_{i=1}^{M}\biggl(\sum_{p=0}^{n}a_{p}z_{i}^{p}\biggr)
 \frac{\del^{2}}{\del z_{i}^{2}},
\end{align}
where $a_{p}$ are constants. Then, the next task is to examine which of
$a_{p}$ can be non-zero free parameters. From a \emph{dimensional analysis},
we easily see that the each term in the r.h.s. of Eq.~\eqref{eq:2ndop2} for
a fixed $p$ must be expressed in terms of $\sigma$ as
\begin{align}
\label{eq:2ndop3}
\sum_{i=1}^{M}z_{i}^{p}\frac{\del^{2}}{\del z_{i}^{2}}
 =\sum_{k,l=1}^{M}\sum_{\{\bm_{i}\}_{k+l+p-2}}\!\!\!\!
 A_{\{\bm_{i}\}_{k+l+p-2},kl}^{[p]}\prod_{i=1}^{M}\sigma_{i}^{\bm_{i}}
 \frac{\del^{2}}{\del\sigma_{k}\del\sigma_{l}},
\end{align}
with a suitable set of constants $A_{\{\bm_{i}\}_{k+l+p-2},kl}^{[p]}$.
On the other hand, we can read from Eqs.~\eqref{eq:polA0}--\eqref{eq:polAkl}
that the l.h.s. of Eq.~\eqref{eq:2ndop2} contains operators
of the form
\begin{align}
\label{eq:f2nd}
\prod_{i=1}^{M}\sigma_{i}^{m_{i}}
 \frac{\del^{2}}{\del\sigma_{k}\del\sigma_{l}},
\end{align}
only for $0\leq\sum_{i=1}^{M}im_{i}\leq k+l+2$. From this fact and
Eq.~\eqref{eq:2ndop3}, we readily know that Eq.~\eqref{eq:2ndop2} cannot
be satisfied unless $a_{p}=0$ for $p>4$. From Eq.~\eqref{eq:polAkl},
we see that the l.h.s. of Eq.~\eqref{eq:2ndop2} contains \emph{all}
the set of operators of the form \eqref{eq:f2nd} for
$(0\leq)k+l-2\leq\sum_{i=1}^{M}im_{i}\leq k+l$, and hence
Eq.~\eqref{eq:2ndop2} can be satisfied for all non-zero values of
$a_{0}$, $a_{1}$, and $a_{2}$, with suitable values
of the parameters $A_{\{m_{i}\}_{k+l},kl}$ in $\bA_{kl}(\sigma)$.
However, only the second term in the r.h.s. of Eq.~\eqref{eq:polAkl}
produces the operators of the form \eqref{eq:f2nd} for
$\sum_{i=1}^{M}im_{i}=k+l+1$ and is insufficient to express the operator
in the r.h.s. of Eq.~\eqref{eq:2ndop2} for $p=3$, cf. Section~B.5
in Ref.~\cite{Ta04}. Hence, Eq.~\eqref{eq:2ndop2} cannot be satisfied
unless $a_{3}=0$ and $A_{10,\{\bm_{i}\}_{k},k}=0$. A similar observation
for $\sum_{i=1}^{M}=k+l+2$ leads to $a_{4}=0$ and $A_{10,10}=A_{20,00}=0$.
Summarizing the analyses so far, we conclude that $a_{p}=0$
for all $p>2$ and thus the coefficient of the second-order operator of
the type A$'$ gauged Hamiltonians of the form \eqref{eq:gaugH} must be
\begin{align}
\label{eq:funAz}
A(z_{i})=a_{2}z_{i}^{2}+a_{1}z_{i}+a_{0}.
\end{align}
Simultaneously, all the following parameters in $\bA_{0}(\sigma)$ and
$\bA_{kl}(\sigma)$ must vanish:
\begin{align}
\label{eq:cond6}
A_{10,\{\bm_{i}\}_{k},k}=A_{10,10}=A_{20,00}=0.
\end{align}

Next, we shall examine the part of the first-order operators. With the aid
of Eq.~\eqref{eq:chain}, the part of the first-order differential operators
in the general type A$'$ operators \eqref{eq:tA'op2} reads
\begin{align}
\label{eq:1stop1}
\sum_{k=1}^{M}\bigl[\bB_{0}(\sigma)k\sigma_{k}-\bB_{k}(\sigma)\bigr]
 \frac{\del}{\del\sigma_{k}}=\sum_{i=1}^{M}S_{i}(z)\frac{\del}{\del z_{i}},
\end{align}
with
\begin{align}
\label{eq:Si}
S_{i}(z)
=\frac{\sum_{k=1}^{M}(-1)^{k+1}z_{i}^{M-k}\bigl[\bB_{0}(\sigma)
 k\sigma_{k}-\bB_{k}(\sigma)\bigr]}{\prod_{j(\neq i)}^{M}(z_{i}-z_{j})}
\equiv\frac{\cS_{i}(z)}{\prod_{j(\neq i)}^{M}(z_{i}-z_{j})}.
\end{align}
Thus, the derivative of $S_{i}(z)$ with respect to $z_{l}(l\neq i)$
in our case is calculated as
\begin{align}
\label{eq:dSi}
\frac{\del S_{i}(z)}{\del z_{l}}
 =\frac{(z_{i}-z_{l})\del_{l}\cS_{i}(z)+\cS_{i}(z)}{(z_{i}-z_{l})^{2}
 \prod_{j(\neq i,l)}^{M}(z_{i}-z_{j})}.
\end{align}
Therefore, the integrability condition \eqref{eq:cond5} can be satisfied
if and only if the numerator of the r.h.s. of Eq.~\eqref{eq:dSi} has
the following form:
\begin{align}
\label{eq:cond7}
(z_{i}-z_{l})\del_{l}\cS_{i}(z)+\cS_{i}(z)=f_{1}(z_{i},z_{l})
 f_{2}(\sigma)A(z_{i})\prod_{j(\neq i,l)}^{M}(z_{i}-z_{j}),
\end{align}
where $f_{1}(z_{i},z_{l})=f_{1}(z_{l},z_{i})$ and $f_{2}(\sigma)$ is
a function depending solely on the elementary symmetric polynomials.

On the other hand, under the condition \eqref{eq:cond6} satisfied,
the coefficient of the first-order
differential operators in Eq.~\eqref{eq:tA'op2} reads
\begin{align}
\label{eq:co1st}
\bB_{0}(\sigma)k\sigma_{k}-\bB_{k}(\sigma)=-B_{10}k\sigma_{1}\sigma_{k}
 +\sum_{\{m_{i}\}_{k}}B_{\{m_{i}\}_{k},k}\prod_{i=1}^{M}\sigma_{i}^{m_{i}}.
\end{align}
Hence, $\cS_{i}(z)$ defined by Eq.~\eqref{eq:Si} is a polynomial of degree
at most $M+1$ in the variables $z$. From Eq.~\eqref{eq:cond7},
we conclude that the combination of the functions $f_{1}(z_{i},z_{l})
f_{2}(\sigma)A(z_{i})$ in the r.h.s. of Eq.~\eqref{eq:cond7} must be
a polynomial of degree at most 3 in the variables $z$. Let us first
examine the highest-degree term. It comes from the first term in the r.h.s.
of Eq.~\eqref{eq:co1st}. The corresponding term in $\cS_{i}(z)$ is
\begin{align}
\cS_{i}(z)=-B_{10}\sigma_{1}\sum_{k=1}^{M}(-1)^{k+1}k z_{i}^{M-k}\sigma_{k}.
\end{align}
With the aid of the formulas \eqref{eq:form1}--\eqref{eq:form3}, we have
\begin{align}
\label{eq:hdterm1}
(z_{i}-z_{l})\del_{l}\cS_{i}(z)+\cS_{i}(z)=-B_{10}z_{i}(z_{i}-z_{l})^{2}
 \prod_{j(\neq i,l)}^{M}(z_{i}-z_{j}).
\end{align}
Comparing with Eq.~\eqref{eq:cond7}, we see that
$f_{1}(z_{i},z_{l})=(z_{i}-z_{l})^{2}$ and $f_{2}(\sigma)=\text{const.}$
for the highest-degree term. However, the remaining term $-B_{10}z_{i}$
is a monomial of first-degree. This means that the highest-degree term
(together with lower-degree terms) cannot be expressed as
Eq.~\eqref{eq:cond7} unless $A(z_{i})$ is a polynomial of first-degree.
In other words, the highest-degree term can exist if and only if $a_{2}=0$
in Eq.~\eqref{eq:funAz}. Hence, the possible form of the formula which
includes the highest-degree term must be
\begin{align}
\label{eq:hdterm2}
(z_{i}-z_{l})\del_{l}\cS_{i}(z)+\cS_{i}(z)=-2gA(z_{i})(z_{i}-z_{l})^{2}
 \delta_{a_{2},0}\prod_{j(\neq i,l)}^{M}(z_{i}-z_{j}),
\end{align}
where $g$ is a constant. It is evident that the term proportional to
$a_{1}$ in the r.h.s. of Eq.~\eqref{eq:hdterm2} comes from the
highest-degree term given by Eq.~\eqref{eq:hdterm1} and thus the constant
$B_{10}$ in Eq.~\eqref{eq:hdterm1} is expressed as
\begin{align}
\label{eq:B10}
B_{10}=2ga_{1}\delta_{a_{2},0}.
\end{align}
We will later see that there in fact exists the term in $S_{i}(z)$ which
corresponds to the term proportional to $a_{0}$ in the r.h.s. of
Eq.\eqref{eq:hdterm2}. Next, we shall examine the lower-degree terms,
namely, the terms of degree less than $M+1$ in $\cS_{i}(z)$ which result
in the terms of degree less than 3 in the combination
$f_{1}(z_{i},z_{l})f_{2}(\sigma)A(z_{i})$. As we have already obtained
the terms which can only exist in the case of $a_{2}=0$, we can assume
here that $a_{2}\neq 0$ and thus $A(z_{i})$ is a strictly second-degree
polynomial. Hence, only the possible form for the lower-degree
terms is $f_{1}(z_{i},z_{l})f_{2}(\sigma)=-2c$, $c$ is a constant,
and we have 
\begin{align}
\label{eq:ldterm}
(z_{i}-z_{l})\del_{l}\cS_{i}(z)+\cS_{i}(z)=-2cA(z_{i})
 \prod_{j(\neq i,l)}^{M}(z_{i}-z_{j}).
\end{align}

Substituting the higher-degree term \eqref{eq:hdterm2} together with
the lower-degree term \eqref{eq:ldterm} in Eq.~\eqref{eq:dSi},
we eventually obtain 
\begin{align}
\frac{\del S_{i}(z)}{\del z_{l}}=-2gA(z_{i})\delta_{a_{2},0}
 -2c\frac{A(z_{i})}{(z_{i}-z_{l})^{2}}\quad(i\neq l).
\end{align}
This set of differential equations can be easily integrated as
\begin{align}
\label{eq:SiA'}
S_{i}(z)=-Q(z_{i})-2g\sigma_{1}A(z_{i})\delta_{a_{2},0}
 -2c\sum_{j(\neq i)}^{M}\frac{A(z_{i})}{z_{i}-z_{j}},
\end{align}
where $Q(z_{i})$ is a function of a single variable. This term should
come from at most $M$th-degree terms in $\cS_{i}(z)$ which cancel
the denominator of the last term in Eq.~\eqref{eq:Si} so that it depends
solely on the single variable $z_{i}$. The denominator is of degree $M-1$
and thus $Q(z_{i})$ must be a polynomial of at most first-degree:
\begin{align}
\label{eq:funQz}
Q(z_{i})=b_{1}z_{i}+b_{0}.
\end{align}
Next, we shall check whether Eq.~\eqref{eq:1stop1} actually holds for
Eqs.~\eqref{eq:co1st} and \eqref{eq:SiA'}. From the formulas
\eqref{eqs:formBC}, the r.h.s. of Eq.~\eqref{eq:1stop1}
with $S_{i}(z)$ given by Eq.~\eqref{eq:SiA'} is expressed as
\begin{multline}
\sum_{i=1}^{M}S_{i}(z)\frac{\del}{\del z_{i}}
 =\sum_{k=1}^{M}\Bigl\{(M-k+1)(M-k+2)ca_{0}\sigma_{k-2}\\
-(M-k+1)\bigl[b_{0}+(M-k)ca_{1}\bigr]\sigma_{k-1}
 -k\bigl[b_{1}+(2M-k-1)ca_{2}\bigr]\sigma_{k}\\
-2(M-k+1)ga_{0}\delta_{a_{2},0}\sigma_{1}\sigma_{k-1}
 -2a_{1}g\delta_{a_{2},0}k\sigma_{1}\sigma_{k}\Bigr\}
 \frac{\del}{\del\sigma_{k}}.
\end{multline}
We now easily see that all the terms in the braces in the r.h.s. of
the above equation are contained in Eq.~\eqref{eq:co1st} for each fixed
$k$. Hence, all the term in Eq.~\eqref{eq:SiA'} can indeed exist.
Substituting Eq.~\eqref{eq:SiA'} in Eq.~\eqref{eq:cond3} and using
Eq.~\eqref{eq:cond4}, we have
\begin{align}
\label{eq:dW}
\frac{\del\cW}{\del z_{i}}=-\frac{Q(z_{i})}{2A(z_{i})}
 +\frac{A'(z_{i})}{4A(z_{i})}-g\sigma_{1}\delta_{a_{2},0}
 -c\sum_{j(\neq i)}^{M}\frac{1}{z_{i}-z_{j}}.
\end{align}
Again, we can easily integrate the above set of differential equations
to obtain
\begin{align}
\label{eq:tA'g1}
\cW=&\,-\sum_{i=1}^{M}\int\dd z_{i}\,\frac{Q(z_{i})}{2A(z_{i})}
 +\frac{1}{4}\sum_{i=1}^{M}\ln\bigl|A(z_{i})\bigr|
 -\frac{g}{2}\sigma_{1}^{2}\delta_{a_{2},0}\notag\\
&\,-c\sum_{i<j}^{M}\ln |z_{i}-z_{j}|,
\end{align}
where we have omitted the integral constant.

Finally from Eqs.~\eqref{eq:gaugH} and \eqref{eq:SiA'}, we find that
the gauged Hamiltonians $\tH_{\cN}^{(A')}$ which preserve the type A$'$
space \eqref{eq:mtypA'} must have the following form:
\begin{align}
\label{eq:gA'Ham}
\tH_{\cN}^{(A')}=-\sum_{i=1}^{M}A(z_{i})\frac{\del^{2}}{\del z_{i}^{2}}
 -\sum_{i=1}^{M}B_{i}(z)\frac{\del}{\del z_{i}}-2c\sum_{i\neq j}^{M}
 \frac{A(z_{i})}{z_{i}-z_{j}}\frac{\del}{\del z_{i}}
 -\bC(\sigma),
\end{align}
where $A(z_{i})$ and $B_{i}(z)$ are given by Eq.~\eqref{eq:funAz} and by
\begin{align}
\label{eq:funBz}
B_{i}(z)=2g\sigma_{1}A(z_{i})\delta_{a_{2},0}+Q(z_{i}),
\end{align}
respectively.
{}From Eqs.~\eqref{eq:polC}, \eqref{eq:cond6}, and \eqref{eq:B10},
the function $\bC$ in Eq.~\eqref{eq:gA'Ham} is calculated as
\begin{align}
\label{eq:funCz}
\bC(\sigma)=-2(\cN-1)ga_{1}\sigma_{1}\delta_{a_{2},0}+c_{0}.
\end{align}
It is easily shown that the gauged Hamiltonian \eqref{eq:gA'Ham} can be
actually cast into a Schr\"odinger operator by a gauge transformation
\begin{align}
\label{eq:tA'Ham}
H_{\cN}^{(A')}&=\ee^{-\cW}\tH_{\cN}^{(A')}\ee^{\cW}\notag\\
&=-\frac{1}{2}\sum_{i=1}^{M}\frac{\del^{2}}{\del q_{i}^{2}}
 +\frac{1}{2}\sum_{i=1}^{M}\left[\left(\frac{\del\cW}{\del q_{i}}\right)^{2}
 -\frac{\del^{2}\cW}{\del q_{i}^{2}}\right]-\bC(\sigma),
\end{align}
if the gauge potential $\cW$ is chosen as Eq.~\eqref{eq:tA'g1}
and the function $z(q)$ which determines the change of variables satisfies
\begin{align}
\label{eq:zqdef}
z'(q)^{2}=2A\bigl(z(q)\bigr)=2\bigl(a_{2}z(q)^{2}+a_{1}z(q)+a_{0}\bigr).
\end{align}
In this case, Eqs.~\eqref{eq:solv1}, \eqref{eq:cond6}, and \eqref{eq:B10}
tell us that the gauged Hamiltonian \eqref{eq:gA'Ham} is not only
quasi-solvable but also solvable if and only if
\begin{align}
\label{eq:solv2}
B_{10}=ga_{1}\delta_{a_{2},0}=0.
\end{align}

It is apparent from the construction that the Hamiltonian \eqref{eq:tA'Ham}
preserves the space $\cV_{\cN;M}^{(A')}$ defined by
\begin{align}
\label{eq:invsp}
\cV_{\cN;M}^{(A')}=\ee^{-\cW}\,\tcV_{\cN;M}^{(A')}.
\end{align}
Hence, the type A$'$ Hamiltonian $H_{\cN}^{(A')}$ can be (locally)
diagonalized in the finite dimensional space \eqref{eq:invsp}. We shall
thus call the space \eqref{eq:invsp} the \emph{solvable sector} of
$H_{\cN}^{(A')}$.

\section{Classification of the Models}
\label{sec:class}

We shall now explicitly compute the concrete form of the type A$'$
quantum Hamiltonians. From Eqs.~\eqref{eq:funAz}, \eqref{eq:dW}, and
\eqref{eq:funCz}, the potential term in Eq.~\eqref{eq:tA'Ham} is
explicitly calculated in terms of $z$ as
\begin{align}
\label{eq:tA'pot}
V=&\,\sum_{i=1}^{M}\frac{1}{16A(z_{i})}\bigl[ 2Q(z_{i})-A'(z_{i})\bigr]
 \bigl[ 2Q(z_{i})-3A'(z_{i})\bigr]\notag\\
&\,+g\biggl[\sigma_{1}(z)\sum_{i=1}^{M}Q(z_{i})
 +\bigl(g\sigma_{1}(z)^{2}+1\bigr)\sum_{i=1}^{M}A(z_{i})+a_{1}(M,\cN)
 \sum_{i=1}^{M}z_{i}\biggr]\delta_{a_{2},0}\notag\\
&\,+c(c-1)\sum_{i<j}^{M}\frac{A(z_{i})+A(z_{j})}{(z_{i}-z_{j})^{2}}+V_{0},
\end{align}
where, and in what follows, $V_{0}$ denotes an arbitrary constant, and
the coupling constant $a_{1}(M,\cN)$ is given by,
\begin{align}
a_{1}(M,\cN)=\bigl[2(\cN-1)+M(M-1)c\bigr] a_{1}.
\end{align}
{}From Eq.~\eqref{eq:zqdef}, the change of variable is determined by
the following integral:
\begin{align}
\label{eq:chang}
\pm(q-q_0)=\int\frac{\dd z}{\sqrt{2(a_{2}z^{2}+a_{2}z+a_{1})}}\,.
\end{align}
In contrast to the type A case where the systems are constructed from
the $\fsl(M+1)$ generators, our present models do not have full
$GL(2,\bbR)$ invariance as has been mentioned previously in Section
\ref{sec:gentA}. However, the remaining invariance under the dilatation
and translation enables us to classify the type A$'$ models. Indeed,
it is readily shown that $A(z)$ can be cast into one of the canonical forms
listed in Table~\ref{tb:canon} by a combination of the dilatation and
translation.
\begin{table}[h]
\caption{Canonical forms for the polynomial $A(z)$, \eqref{eq:funAz},
where $\nu$ is a positive real constant.}
\label{tb:canon}
\begin{center}
\[
\begin{tabular}{ll}
\hline
Case\qquad & Canonical Form\\
\hline
I   & $1/2$\\
II  & $2z$\\
III & $\pm 2\nu z^{2}$\\ 
IV  & $\pm 2\nu (z^{2}-1)$\\
IV$'$ & $\pm 2\nu (z^{2}+1)$\\
\hline 
\end{tabular}
\]
\end{center}
\end{table}

Furthermore, we note that from Eq.~\eqref{eq:zqdef}, a rescaling of
the coefficients $a_{i}$, $b_{i}$, $c_{0}$ by an overall constant factor
$\nu$ has the following effect on the change of variable $z(q)$:
\begin{align}
\label{eq:zscale}
z(q\,;\nu a_{i},\nu b_{i},\nu c_{0})=z(\rnu\,q\,;a_{i},b_{i},c_{0}).
\end{align}
{}From this equation and Eqs.~\eqref{eq:funAz}, \eqref{eq:funQz},
\eqref{eq:tA'g1}, and \eqref{eq:tA'pot}, we easily obtain the identities
\begin{subequations}
\label{eqs:EFWV}
\begin{align}
\label{eq:cWscale}
\cW(q\,;\nu a_{i},\nu b_{i},\nu c_{0})&=\cW(\rnu\,q\,;a_{i},b_{i},c_{0}),\\
\label{eq:Vscale}
V(q\,;\nu a_{i},\nu b_{i},\nu c_{0})&=\nu\,V(\rnu\,q\,;a_{i},b_{i},c_{0}).
\end{align}
\end{subequations}
We shall therefore set $\nu=1$ in the canonical forms in Cases
II--IV${}^{(')}$, the models corresponding to an arbitrary value of
$\nu$ following easily from Eqs.~\eqref{eq:zscale} and \eqref{eqs:EFWV}.
It should also be obvious from Eq.~\eqref{eq:chang} that the change of
variable $z(q)$, and hence the potential $V$ determining each model,
are defined up to the transformation $q\mapsto\pm(q-q_{0})$, where
$q_{0}\in\bbR$ is a constant.
The solvability condition \eqref{eq:solv2} implies that except for
the model with $g\neq 0$ corresponding to Case II in Table~\ref{tb:canon}
all the obtainable models are not only quasi-solvable but also solvable.

As we will see below, the potentials in Case I, III, and IV$'$ have
singularities at $q_{i}=q_{j}$ for all $i\neq j$ in the subspace
$\{-\Omega\leq q_{i}<\Omega\}\in\bbR^{M}$ where $\Omega$ is a submultiple
of the real period of the potential or $\Omega=\infty$ when the potential
is non-periodic. Similarly, the potentials in Case II and IV are singular
at $q_{i}=0$ and $q_{i}=\pm q_{j}$ for all $i\neq j$ in the same subspace
of $\bbR^{M}$. Hence, the Hamiltonians are naturally defined on
\begin{align}
\label{eq:Weyls}
0<q_{M}<\dots<q_{1}<\Omega.
\end{align}
Normalizability of the solvable sector \eqref{eq:invsp} on the space
\eqref{eq:Weyls} depends on the behavior of the gauge potential $\cW(q)$
in the each case.
The finiteness of the $L^{2}$ norm of the two-body wave function in
the solvable sector $\cV_{\cN;M}^{(A')}$ in general leads $c>-1/2$, where
$c$ denotes the coupling constant of the two-body
interaction appeared in the last line of Eq.~\eqref{eq:tA'pot}.

\subsection{Case I: $A(z)=1/2$}

\emph{Change of variable}:\quad $z(q)=q$.\\

\emph{Potential}:
\begin{align}
\label{eq:pot1}
V(q)=&\,g\biggl(\frac{M}{2}g+b_{1}\biggr)\biggl(\sum_{i=1}^{M}q_{i}\biggr)^{2}
 +Mb_{0}g\sum_{i=1}^{M}q_{i}+\frac{1}{2}\sum_{i=1}^{M}(b_{1}q_{i}+b_{0})^{2}
 \notag\\
&\,+c(c-1)\sum_{i<j}^{M}\frac{1}{(q_{i}-q_{j})^2}+V_{0}.
\end{align}
\emph{Gauge potential}:
\begin{align}
\label{eq:gpot1}
\cW(q)=-\frac{g}{2}\biggl(\sum_{i=1}^{M}q_{i}\biggr)^{2}
 -\frac{b_{1}}{2}\sum_{i=1}^{M}q_{i}^{2}-b_{0}\sum_{i=1}^{M}q_{i}
 -c\sum_{i<j}^{M}\ln|q_{i}-q_{j}|.
\end{align}
When $g=0$, this case corresponds to the rational $A_{M-1}$ type
Calogero--Sutherland model \cite{Ca71} and is identical with the solvable
model corresponding to Case I of the type A models, Eqs.~(7.8)--(7.9) with
$b_{2}=0$ in Ref.~\cite{Ta04}. Hence, the above model provides an example
of deformed Calogero--Sutherland models which preserve quantum solvability.
The parameter $b_{0}$ corresponds to the translational degree of freedom
and is irrelevant. In fact, if we employ the translational freedom
mentioned earlier below Eqs.~\eqref{eqs:EFWV} and apply the translation
$q_{i}\mapsto q_{i}-b_{0}/(b_{1}+Mg)$ in
Eqs.~\eqref{eq:pot1}--\eqref{eq:gpot1}, the potential and gauge potential
become
\begin{align}
V(q)&=g\biggl(\frac{M}{2}g+b_{1}\biggr)\biggl(\sum_{i=1}^{M}q_{i}\biggr)^{2}
 +\frac{b_{1}^{2}}{2}\sum_{i=1}^{M}q_{i}^{2}+c(c-1)\sum_{i<j}^{M}
 \frac{1}{(q_{i}-q_{j})^{2}}+V_{0},\\
\cW(q)&=-\frac{g}{2}\biggl(\sum_{i=1}^{M}q_{i}\biggr)^{2}
 -\frac{b_{1}}{2}\sum_{i=1}^{M}q_{i}^{2}-c\sum_{i<j}^{M}\ln|q_{i}-q_{j}|,
\end{align}
and have no dependence on $b_{0}$ any more. It is now obvious that the model
is exactly the rational $A_{M-1}$ type Calogero--Sutherland model with the
center-of-mass coordinate subjected to the harmonic oscillator potential.
Since we can easily separate the center-of-mass coordinate from the others,
quantum solvability of the above model is readily understood.

The one-body part of the potential has no singularities and hence a natural
choice is $\Omega=\infty$. In this choice, the form of the gauge potential
\eqref{eq:gpot1} tells us that the solvable sector
\eqref{eq:invsp} is square integrable on the space \eqref{eq:Weyls}
as long as $g<0$, $b_{1}<0$, and $c>-1/2$. Hence, the model \eqref{eq:pot1}
is exactly solvable on Eq.~\eqref{eq:Weyls} in these parameter regions.

\subsection{Case II: $A(z)=2z$}

\emph{Change of variable}:\quad $z(q)=q^{2}$.\\

\emph{Potential}:
\begin{align}
\label{eq:pot2}
V(q)=&\,\biggl(\frac{b_{1}^{2}}{8}+2(2\cN-1)g+Mb_{0}g+2M(M-1)cg\biggr)
 \sum_{i=1}^{M}q_{i}^{2}\notag\\
&\,+2g^{2}\biggl(\sum_{i=1}^{M}q_{i}^{2}\biggr)^{3}
 +b_{1}g\biggl(\sum_{i=1}^{M}q_{i}^{2}\biggr)^{2}
 +\frac{(b_{0}-1)(b_{0}-3)}{8}\sum_{i=1}^{M}\frac{1}{q_{i}^{2}}\notag\\
&\,+c(c-1)\sum_{i<j}^{M}\biggl[\frac{1}{(q_{i}-q_{j})^{2}}
 +\frac{1}{(q_{i}+q_{j})^{2}}\biggr]+V_{0}.
\end{align}
\emph{Gauge potential}:
\begin{align}
\label{eq:gpot2}
\cW(q)=-\frac{g}{2}\biggl(\sum_{i=1}^{M}q_{i}^{2}\biggr)^{2}
 -\frac{b_{1}}{4}\sum_{i=1}^{M}q_{i}^{2}
 -\frac{b_{0}-1}{2}\sum_{i=1}^{M}\ln|q_{i}|
 -c\sum_{i<j}^{M}\ln\bigl|q_{i}^{2}-q_{j}^{2}\bigr|.
\end{align}
This case corresponds to the rational $BC_{M}$ type Calogero--Sutherland
model \cite{OP83} when $g=0$. Furthermore, the above model is quasi-solvable
when $g\neq 0$. To the best of our knowledge, it is new and provides
another quasi-solvable deformation of Calogero--Sutherland models which is
different from the Inozemtsev type classified in Ref.~\cite{Ta04}.
In this respect, we should refer to similar quasi-solvable models in
the literature, namely, Eq.~(23) in Ref.~\cite{MRT96} and Eq.~(3.7)
in Ref.~\cite{KM98}. These models are deformations of the rational
$A_{M-1}$ type Calogero--Sutherland system and thus their
two-body interaction terms are different from our $BC_{M}$ type.
Furthermore, the solvable sectors of these models are spanned by
monomials of a \emph{single} variable while that of our present model
is by monomials of $M$ variables \eqref{eq:mtypA'}.

The one-body part of the potential is only singular at $q_{i}=0$ and hence
a natural choice is $\Omega=\infty$. In this choice, we see from
Eq.~\eqref{eq:gpot2} that the solvable sector \eqref{eq:invsp} is
square integrable on the space \eqref{eq:Weyls} as long as $g<0$ and
$c>-1/2$. Hence, the model \eqref{eq:pot2} is quasi-exactly solvable
on the space \eqref{eq:Weyls} in these parameter regions.
When $M=1$, the above model becomes
\begin{align}
V(q)=\frac{1}{8}q^{2}(4gq^{2}+b_{1})^{2}+(4\cN+b_{0}-2)gq^{2}
 +\frac{(b_{0}-1)(b_{0}-3)}{8q^{2}}+V_{0},
\end{align}
and thus exactly reduces to the well-known one-body quasi-solvable sextic
anharmonic oscillator, classified in Case II of the type A models,
Eqs.~(7.11)--(7.12) with $b_{2}=4g$ and $c_{1}=(b_{0}-1)/2$ in
Ref.~\cite{Ta04}.

\subsection{Case IIIa: $A(z)=2z^{2}$}

\emph{Change of variable}:\quad $z(q)=\ee^{2q}$.\\

\emph{Potential}:
\begin{align}
\label{eq:pot3}
V(q)=\frac{b_{0}(b_{1}-4)}{4}\sum_{i=1}^{M}\ee^{-2q_{i}}
 +\frac{b_{0}^{2}}{8}\sum_{i=1}^{M}\ee^{-4q_{i}}
 +c(c-1)\sum_{i<j}^{M}\frac{1}{\sinh^{2}(q_{i}-q_{j})}+V_{0}.
\end{align}
\emph{Gauge potential}:
\begin{align}
\label{eq:gpot3}
\cW(q)=&\,\frac{b_{0}}{4}\sum_{i=1}^{M}\ee^{-2q_{i}}
 -\Bigl(\frac{b_{1}}{2}-1+(M-1)c\Bigr)\sum_{i=1}^{M}q_{i}\notag\\
&\,-c\sum_{i<j}^{M}\ln\bigl|\sinh(q_{i}-q_{j})\bigr|.
\end{align}
This model is the hyperbolic $A_{M-1}$ Calogero--Sutherland model \cite{Su71}
in the external Morse potential and completely the same as the model
corresponding to Case III of the type A models, Eqs.~(7.15)--(7.16) with
$b_{2}=0$ in Ref.~\cite{Ta04}.

\subsection{Case IIIb: $A(z)=-2z^{2}$}

The formulas of the potential and gauge potential for this case can
be easily reduced from Eqs.~\eqref{eq:pot3}--\eqref{eq:gpot3} using
Eqs.~\eqref{eqs:EFWV} with $\nu=-1$. The change of variable is
$z(q)=\ee^{2\ii q}$.

\subsection{Case IVa: $A(z)=2(z^{2}-1)$}

\emph{Change of variable}:\quad $z(q)=\cosh 2q$.\\

\emph{Potential}:
\begin{align}
\label{eq:pot4}
V(q)=&\,\frac{(b_{1}-b_{0}-2)(b_{1}-b_{0}-6)}{8}\sum_{i=1}^{M}
 \frac{1}{\sinh^{2}2q_{i}}+\frac{b_{0}(b_{1}-4)}{8}\sum_{i=1}^{M}
 \frac{1}{\sinh^{2}q_{i}}\notag\\
&\,+c(c-1)\sum_{i<j}^{M}\biggl[\frac{1}{\sinh^{2}(q_{i}-q_{j})}
 +\frac{1}{\sinh^{2}(q_{i}+q_{j})}\biggr]+ V_{0}.
\end{align}
\emph{Gauge potential}:
\begin{align}
\label{eq:gpot4}
\cW(q)=&\,-\frac{b_{1}-2}{4}\sum_{i=1}^{M}\ln|\sinh 2q_{i}|
 -\frac{b_{0}}{4}\sum_{i=1}^{M}\ln|\tanh q_{i}|\notag\\
&\,-c\sum_{i<j}^{M}\ln\bigl|\sinh(q_{i}-q_{j})\sinh(q_{i}+q_{j})\bigr|.
\end{align}
This model is the hyperbolic $BC_{M}$ Calogero--Sutherland model \cite{OP83}
and completely the same as the model corresponding to Case IV of the type A
models, Eqs.~(7.22)--(7.23) with $b_{2}=0$ in Ref.~\cite{Ta04}.

\subsection{Case IVb: $A(z)=-2(z^{2}-1)$}

The formulas of the potential and gauge potential for this case can
be easily reduced from Eqs.~\eqref{eq:pot4}--\eqref{eq:gpot4} using
Eqs.~\eqref{eqs:EFWV} with $\nu=-1$. The change of variable is
$z(q)=\cos 2q$.

\subsection{Case IV$\,'$a: $A(z)=2(z^{2}+1)$}

\emph{Change of variable}:\quad $z=\sinh 2q$.\\

\emph{Potential}:
\begin{align}
\label{eq:pot4'}
V(q)=&\,\frac{b_{0}^{2}-(b_{1}-2)(b_{1}-6)}{8}\sum_{i=1}^{M}
 \frac{1}{\cosh^{2}2q_{i}}+\frac{b_{0}(b_{1}-4)}{4}\sum_{i=1}^{M}
 \frac{\sinh 2q_{i}}{\cosh^{2}2q_{i}}\notag\\
&\,+c(c-1)\sum_{i<j}^{M}\biggl[\frac{1}{\sinh^{2}(q_{i}-q_{j})}
 -\frac{1}{\cosh^{2}(q_{i}+q_{j})}\biggr]+ V_{0}.
\end{align}
\emph{Gauge potential}:
\begin{align}
\label{eq:gpot4'}
\cW(q)=&\,-\frac{b_{1}-2}{4}\sum_{i=1}^{M}\ln |\cosh 2q_{i}|
 -\frac{b_{0}}{4}\sum_{i=1}^{M}\gd 2q_{i}\notag\\
&\,-c\sum_{i<j}^{M}\bigl|\sinh(q_{i}-q_{j})\cosh(q_{i}+q_{j})\bigr|,
\end{align}
where $\gd q=\arctan(\sinh q)$ is the Gudermann function.
This model is another hyperbolic $BC_{M}$ Calogero--Sutherland model
and completely the same as the model corresponding to Case IV$'$ of the
type A models, Eqs.~(7.26)--(7.27) with $b_{2}=0$ in Ref.~\cite{Ta04}.

\subsection{Case IV$\,'$b: $A(z)=-2(z^{2}+1)$}

The formulas of the potential and gauge potential for this case can
be easily reduced from Eqs.~\eqref{eq:pot4'}--\eqref{eq:gpot4'} using
Eqs.~\eqref{eqs:EFWV} with $\nu=-1$. The change of variable is
$z(q)=\ii\sin 2q$.

\section{A New Type C Generalization Based on the Type A$'$ Scheme}
\label{sec:newtC}

In the preceding sections, we have investigated (at most) second-order
linear differential operators preserving the type A$'$ space
\eqref{eq:mtypA'}. This new generalization of the single-variable type
A space to several variables suggests a new multivariate generalization
of the single-variable type C space. The type C monomial space of a single
variable $z$ is defined by \cite{GT05}
\begin{align}
\label{eq:stypC}
\tcV_{\cN_{1}\!,\cN_{2}}^{(C)}=\tcV_{\cN_{1}}^{(A)}\oplus
 z^{\lambda}\,\tcV_{\cN_{2}}^{(A)},
\end{align}
where $\tcV_{\cN_{i}}^{(A)}$ ($i=1,2$) is a type A monomial space
of dimension $\cN_{i}$ defined by Eq.~\eqref{eq:stypA}, $\cN_{1}$ and
$\cN_{2}$ are positive integers satisfying $\cN_{1}\geq\cN_{2}$, and
$\lambda$ is a real number with the restriction
\begin{align}
\label{eq:restr}
\lambda\in\bbR\setminus\{-\cN_{2},-\cN_{2}+1,\dots,\cN_{1}\},
\end{align}
and with $\lambda\neq -\cN_{2}-1,\,\cN_{1}+1$ if $\cN_{1}=1$ or
$\cN_{2}=1$. In the previous paper~\cite{Ta05a}, we generalized the
type C space of a single variable \eqref{eq:stypC} to several variables
as follows:
\begin{align}
\label{eq:mtypC}
\tcV_{\cN_{1}\!,\cN_{2};M}^{(C)}=\tcV_{\cN_{1};M}^{(A)}\oplus
 \sigma_{M}^{\lambda}\,\tcV_{\cN_{2};M}^{(A)},
\end{align}
where $\tcV_{\cN_{i};M}^{(A)}$ ($i=1,2$) is a multivariate type A
space defined by Eq.~\eqref{eq:mtypA}. The above generalization scheme
based on the type A space \eqref{eq:mtypA} now strongly suggests
another generalization based on the type A$'$ space \eqref{eq:mtypA'}
as the following:
\begin{align}
\label{eq:mtypC'}
\tcV_{\cN_{1}\!,\cN_{2};M}^{(C')}=\tcV_{\cN_{1};M}^{(A')}\oplus
 \sigma_{M}^{\lambda}\,\tcV_{\cN_{2};M}^{(A')}.
\end{align}
Indeed, the latter space also reduces to the single-variable type C space
\eqref{eq:stypC} when $M=1$. Hence, Eq.~\eqref{eq:mtypC'} would provide
a new multivariate generalization of the type C space and we hereafter call
the space \eqref{eq:mtypC'} \emph{type C$\,'$}. In this section, we shall
investigate (at most) second-order linear differential operators which
preserve the type C$'$ space \eqref{eq:mtypC'}.

First of all, the set of linearly independent first-order differential
operators preserving the type C$'$ space is given by,
\begin{align}
\label{eq:1st1C}
F_{\{m_{i}\}_{\bk},\bk}&=\prod_{i=1}^{M}\sigma_{i}^{m_{i}}
 \frac{\del}{\del\sigma_{\bk}}\quad(\bk=1,\dots,M-1),\\
E_{M\!M}&\equiv\sigma_{M}\frac{\del}{\del\sigma_{M}}.
\end{align}
Similarly, the set of the linearly independent second-order differential
operators preserving the type C$'$ space is as follows:
\begin{align}
F_{\{m_{i}\}_{\bk+\bl},\bk\bl}&=\prod_{i=1}^{M}\sigma_{i}^{m_{i}}
 \frac{\del^{2}}{\del\sigma_{\bk}\del\sigma_{\bl}}\quad
 (\bk,\bl=1,\dots,M-1;\,\bk\geq\bl),\\
F_{\{m_{i}\}_{\bk},\bk}E_{M\!M}&=\prod_{i=1}^{M}\sigma_{i}^{m_{i}}
 \sigma_{M}\frac{\del^{2}}{\del\sigma_{M}\del\sigma_{\bk}}
 \quad(\bk=1,\dots,M-1),\\
\label{eq:2nd3C}
F_{\{m_{i}\}_{M}\!,M}(E_{M\!M}-\lambda)&=\prod_{i=1}^{M}\sigma_{i}^{m_{i}}
 \frac{\del}{\del\sigma_{M}}\biggl(\sigma_{M}\frac{\del}{\del\sigma_{M}}
 -\lambda\biggr),\\
\label{eq:2nd4C}
F_{10,00}\equiv\sigma_{1}\biggl(\cN_{1}-1&-\sum_{k=1}^{M}k\sigma_{k}
 \frac{\del}{\del\sigma_{k}}\biggr)\biggl(M\lambda+\cN_{2}-1
 -\sum_{l=1}^{M}l\sigma_{l}\frac{\del}{\del\sigma_{l}}\biggr).
\end{align}
Therefore, the most general quasi-solvable operator of (at most)
second-order which preserves the type C$'$ space \eqref{eq:mtypC'} is
given by the linear combination of all the operators
\eqref{eq:1st1C}--\eqref{eq:2nd4C}:
\begin{align}
\tcH^{(C')}=&\,-\sum_{\bk\geq\bl}^{M-1}\sum_{\{m_{i}\}_{\bk+\bl}}\!\!
 A_{\{m_{i}\}_{\bk+\bl},\bk\bl}F_{\{m_{i}\}_{\bk+\bl},\bk\bl}
 -\sum_{\bk=1}^{M-1}\sum_{\{m_{i}\}_{\bk}}\!\!A_{\{m_{i}\}_{\bk},\bk,M\!M}
 F_{\{m_{i}\}_{\bk},\bk}E_{M\!M}\notag\\
&\,-\sum_{\{m_{i}\}_{M}}\!\!A_{\{m_{i}\}_{M},M\!,M\!M}F_{\{m_{i}\}_{M}\!,M}
 (E_{M\!M}-\lambda)-A_{10,00}F_{10,00}\notag\\
&\,+\sum_{\bk=1}^{M-1}B_{\{m_{i}\}_{\bk},\bk}F_{\{m_{i}\}_{\bk},\bk}
 +B_{M\!M}E_{M\!M}-c_{0},
\end{align}
where again the coefficients $A$ $B$ with indices and $c_{0}$ are real
constants and the summation over the set $\{m_{i}\}_{\bk}$ etc. is
understood to take all the possible set of values $\{m_{1},\dots,m_{M}\}$
indicated in Eq.~\eqref{eq:set1}. In terms of the variables $\sigma$,
the operator $\tcH^{(C')}$ is expressed as
\begin{align}
\label{eq:tC'op2}
\tcH^{(C')}=&\,-\sum_{k,l=1}^{M}\bigl[A_{10,00}kl\sigma_{1}
 \sigma_{k}\sigma_{l}+\bA_{kl}(\sigma)\bigr]
 \frac{\del^{2}}{\del\sigma_{k}\del\sigma_{l}}\notag\\
&\,+\sum_{k=1}^{M}\bigl[A_{10,00}(\cN+M\lambda-k-2)k\sigma_{1}\sigma_{k}
 +\bB_{k}(\sigma)\bigr]\frac{\del}{\del\sigma_{k}}-\bC(\sigma),
\end{align}
where $\bA_{kl}$, $\bB_{k}$, and $\bC$ are polynomials of several variables
given by
\begin{subequations}
\begin{align}
\bA_{\bk\bl}(\sigma)&=\sum_{\{m_{i}\}_{\bk+\bl}}\!\!
  A_{\{m_{i}\}_{\bk+\bl},\bk\bl}\prod_{i=1}^{M}\sigma_{i}^{m_{i}}
 \quad(\bk,\bl=1,\dots,M-1;\,\bk\geq\bl),\\
\bA_{M\!k}(\sigma)&=\sum_{\{m_{i}\}_{k}}A_{\{m_{i}\}_{k},k,M\!M}
 \prod_{i=1}^{M}\sigma_{i}^{m_{i}}\sigma_{M}\quad(k=1,\dots,M),\\
\bB_{\bk}(\sigma)&=-\sum_{\{m_{i}\}_{\bk}}B_{\{m_{i}\}_{\bk},\bk}
 \prod_{i=1}^{M}\sigma_{i}^{m_{i}}\quad(\bk=1,\dots,M-1),\\
\bB_{M}(\sigma)&=-(\lambda-1)\sum_{\{m_{i}\}_{M}}
 A_{\{m_{i}\}_{M}\!,M\!,M\!M}\prod_{i=1}^{M}\sigma_{i}^{m_{i}}
 -B_{M\!M}\sigma_{M},\\[6pt]
\bC(\sigma)&=(\cN_{1}-1)(\cN_{2}+M\lambda-1)A_{10,00}\sigma_{1}+c_{0}.
\end{align}
\end{subequations}
Except for the operator $F_{10,00}$, all the operators in
Eqs.~\eqref{eq:1st1C}--\eqref{eq:2nd3C} leave the type C$'$ space
\eqref{eq:mtypC'} invariant for \emph{arbitrary} natural numbers
$\cN_{1}$ and $\cN_{2}$. Hence, the operator $\tcH^{(C')}$ is not only
quasi-solvable but also solvable if
\begin{align}
A_{10,00}=0.
\end{align}

As in the case of regular type C in Ref.~\cite{Ta05a}, all
the operators in Eqs.~\eqref{eq:1st1C}--\eqref{eq:2nd4C} and hence
the most general type C$'$ operator \eqref{eq:tC'op2} preserve
\emph{separately} both the subspaces of $\tcV_{\cN_{1},\cN_{2};M}^{(C')}$
in Eq.~\eqref{eq:mtypC'}, the fact originally comes from the restriction
\eqref{eq:restr}. In other words, the second-order operators
$\sigma_{M}^{-(k-1)\lambda}\tcH^{(C')}\sigma_{M}^{(k-1)\lambda}$
($k=1,2$) leave the type A$'$ space $\tcV_{\cN_{k}}^{(A')}$ invariant,
respectively. As a consequence, the most general type C$'$ gauged
Hamiltonian $\tH^{(C')}$ must satisfy the following condition:
\begin{align}
\label{eq:conC'}
\tH^{(C')}=\tH_{\cN_{1}}^{(A')}
 =\sigma_{M}^{\lambda}\,\tH_{\cN_{2}}^{(A')}\sigma_{M}^{-\lambda}.
\end{align}
In the above, each of the operators $\tH_{\cN_{k}}^{(A')}$ ($k=1,2$)
is a type A$'$ gauged Hamiltonian and thus has the form
\eqref{eq:gA'Ham} with Eqs.~\eqref{eq:funAz}, \eqref{eq:funQz},
\eqref{eq:funBz}, and \eqref{eq:funCz}. Tracing a completely similar
way to Section 5 in Ref.~\cite{Ta05a} and noting the fact that the
first-order operator $F_{10}$ defined by Eq.~\eqref{eq:1st2} is missing
in the set of the type C$'$ operators \eqref{eq:1st1C}--\eqref{eq:2nd4C},
we find that the most general type C$'$ gauged Hamiltonian satisfying
the condition \eqref{eq:conC'} has the following form:
\begin{align}
\label{eq:gC'Ham}
\tH^{(C')}=-\sum_{i=1}^{M}A(z_{i})\frac{\del^{2}}{\del z_{i}^{2}}
 -\sum_{i=1}^{M}B(z_{i})\frac{\del}{\del z_{i}}-2c\sum_{i\neq j}^{M}
 \frac{A(z_{i})}{z_{i}-z_{j}}\frac{\del}{\del z_{i}}-c_{0},
\end{align}
where
\begin{align}
A(z_{i})&=a_{2}z_{i}^{2}+a_{1}z_{i},\\
B(z_{i})&=b_{1}z_{i}-(\lambda-1)a_{1},
\end{align}
and $a_{i}$, $b_{1}$, $c_{0}$, and $c$ are constants. Comparing
the above results with the most general type C gauged Hamiltonian,
Eqs.~(5.17)--(5.20) in Ref.~\cite{Ta05a}, we see that the type C$'$
gauged Hamiltonian \eqref{eq:gC'Ham} is a special case of the type C
gauged Hamiltonian with $a_{3}=0$. Hence, all the quantum mechanical
models of type C$'$ are included in the ones fully classified in
Ref.~\cite{Ta05a}.

\section{Discussion and Summary}
\label{sec:discus}

\noindent
In this article, we have made a new generalization of the type A
monomial space of a single variable to several variables and have
constructed the most general (at most) second-order quasi-solvable
operator which preserves the new linear space called type A$'$.
Examining the condition under which the type A$'$ second-order
operators can be transformed to Schr\"odinger operators, we have
extracted the most general type A$'$ gauged Hamiltonian. Then, we have
completely classified the type A$'$ quantum Hamiltonians. We have also
investigated a new type C generalization called type C$'$ based on
the type A$'$ space. Combining the results obtained in this article
with the ones in Refs.~\cite{Ta04,Ta05a}, we can summarize the
classification of the type A$'$ quasi-solvable quantum many-body systems
as shown in Table~\ref{tb:class}.

\begin{table}[h]
\caption{Classification of the type A$'$ (quasi-)solvable quantum
many-body systems.}
\label{tb:class}
\begin{center}
\[
\begin{tabular}{lcc}
\hline
Model & Type & Solvable\\
\hline
Rational $A$ Calogero--Sutherland & A$'$, A & $\bigcirc$\\
\quad + quadratic $M$-body interaction & A$'$ & $\bigcirc$\\
Rational $BC$ Calogero--Sutherland & C$'$, C & $\bigcirc$\\
\quad + sextic $M$-body interaction & A$'$ & $\times$\\
Hyp.(Trig.) $A$ Calogero--Sutherland & C$'$, C & $\bigcirc$\\
\quad + external Morse potential & A$'$, A & $\bigcirc$\\
Hyp.(Trig.) $BC$ Calogero--Sutherland & C$'$, C & $\bigcirc$\\
\hline 
\end{tabular}
\]
\end{center}
\end{table}

The meaning of Table~\ref{tb:class} is as follows. For example,
the rational $BC$ Calogero--Sutherland model in the third row
belongs to type C$'$ and C, and it is not only quasi-solvable
but also solvable. If the sextic $M$-body interaction is added
to this model (the fourth row), it belongs to type A$'$ and it is
only quasi-solvable but not solvable. The interpretation for other
models would be straightforward in a similar way.

One may be curious about the fact that some of the models belong to
both type A and A$'$, or to both type C and C$'$. Let us first consider
the former cases. The fact that a model belongs to both type A and A$'$
means that it preserves type A and A$'$ spaces \emph{simultaneously}.
In this respect, we note that for fixed values of $M(>1)$ and $\cN$
the type A$'$ space \eqref{eq:mtypA'} is a subspace of the type A space
\eqref{eq:mtypA}:
\begin{align}
\label{eq:A'subA}
\tcV_{\cN;M}^{(A')}\subset\tcV_{\cN;M}^{(A)}.
\end{align}
However, it does not mean that an operator which preserves a type A space
always preserves a type A$'$ space too nor mean vice versa. For instance,
all the operators of the form
\begin{align}
E_{ij}=\sigma_{i}\frac{\del}{\del\sigma_{j}}\quad(i,j=1,\dots,M;\,i>j),
\end{align}
preserve type A spaces but do not any type A$'$ spaces, while the operator
$F_{10}$ defined by Eq.~\eqref{eq:1st2} preserves the type A$'$ space
with the same $\cN$ as in $F_{10}$ but does not any type A spaces.
{}From the set of the operators \eqref{eqs:1st} and
\eqref{eq:2nd1}--\eqref{eq:2nd4} which preserve the type A$'$ space and
the set of operators (3.10) and (3.12) in Ref.~\cite{Ta04} which preserve
the type A space\footnote{We would like to note that there are typos in
Ref.~\cite{Ta04}; the following operators are missing in Eqs.~(3.12):
\[
\frac{\del^{2}}{\del\sigma_{i}\del\sigma_{j}}=E_{0i}E_{0j},
\]
and Eq.~(3.12c) should be
\[
\sigma_{i}\biggl(\cN-1-\sum_{l=1}^{M}\sigma_{l}\frac{\del}{\del\sigma_{l}}
 \biggr)\sigma_{j}\frac{\del}{\del\sigma_{k}}=E_{i0}E_{jk}.
\]}, we find that the set of linearly independent differential operators of
(at most) second-order which preserve both the type A and A$'$ spaces
simultaneously is the following:
\begin{subequations}
\begin{gather}
\frac{\del}{\del\sigma_{i}},\qquad \sigma_{i}\frac{\del}{\del\sigma_{j}}
 \quad (i\leq j),\\
\frac{\del^{2}}{\del\sigma_{k}\del\sigma_{l}},\quad
 \sigma_{i}\frac{\del^{2}}{\del\sigma_{k}\del\sigma_{l}}\quad
 (i\leq k+l),\quad
\sigma_{i}\sigma_{j}\frac{\del^{2}}{\del\sigma_{k}\del\sigma_{l}}
 \quad (i+j\leq k+l),\\
\label{eq:2nd3CC'}
\sigma_{i}\biggl(\cN-1-\sum_{l=1}^{M}\sigma_{l}\frac{\del}{\del\sigma_{l}}
 \biggr)\sigma_{j}\frac{\del}{\del\sigma_{k}}\quad(i+j\leq k).
\end{gather}
\end{subequations}
The most general quasi-solvable operator of (at most) second-order
$\tcH^{(A,A')}$ is obviously obtained by the linear combination of all
the above operators. In particular, the most general gauged Hamiltonian
$\tH^{(A,A')}$ leaving the type A and A$'$ spaces simultaneously invariant
must be of the form of both the type A gauged Hamiltonian,
Eqs.~(5.12)--(5.14) in Ref.~\cite{Ta04}, and the type A$'$ gauged
Hamiltonian, Eqs.~\eqref{eq:gA'Ham}--\eqref{eq:funCz} with \eqref{eq:funAz}
and \eqref{eq:funQz}. From the observation, we see that the type A$'$ gauged
Hamiltonian \eqref{eq:gA'Ham} belongs to type A too if and only if
$g\delta_{a_{2},0}=0$. We note that if it is the case, the system is
always solvable since the solvability condition \eqref{eq:solv2} is
automatically satisfied in the case. In fact, all the models without
$M$-body interactions in Table~\ref{tb:class} satisfy the condition
$g\delta_{a_{2},0}=0$, thus belong to both type A and A$'$, and are
not only quasi-solvable but also solvable. One of the interesting
consequences is that all the models without $M$-body interactions in
Table~\ref{tb:class} preserve the following infinite flag of finite
dimensional linear spaces:
\begin{equation}
\label{eq:flagA}
\begin{array}{cccccl}
\cV_{1;M}^{(A)} & \subset & \cV_{2;M}^{(A)}
 & \subset\dots\subset & \cV_{\cN;M}^{(A)} & \subset\cdots\\
\parallel & & \cup & & \cup & \\
\cV_{1;M}^{(A')} & \subset & \cV_{2;M}^{(A')}
 & \subset\dots\subset & \cV_{\cN;M}^{(A')} & \subset\cdots,
\end{array}
\end{equation}
where $\cV_{\cN;M}^{(A)}$ are gauge-transformed type A spaces
\begin{align}
\cV_{\cN;M}^{(A)}=\ee^{-\cW}\,\tcV_{\cN;M}^{(A)},
\end{align}
with the same gauge potential $\cW$ as in Eq.~\eqref{eq:invsp}.

The situation in the case of simultaneous type C and C$'$ is
completely analogous to in the case of simultaneous type A and A$'$
discussed just above. The models which belong to both type C and C$'$
preserve the type C and C$'$ spaces simultaneously. From
Eqs.~\eqref{eq:mtypC}, \eqref{eq:mtypC'}, and \eqref{eq:A'subA},
we easily see that for fixed values of $M(>1)$, $\cN_{1}$, and $\cN_{2}$
the type C$'$ space is a subspace of the type C space:
\begin{align}
\tcV_{\cN_{1}\!,\cN_{2};M}^{(C')}\subset\tcV_{\cN_{1}\!,\cN_{2};M}^{(C)}.
\end{align}
The set of linearly independent differential operators of (at most)
second-order which preserve both the type C and C$'$ spaces simultaneously
is the following:
\begin{subequations}
\label{eqs:opsCC'}
\begin{gather}
\frac{\del}{\del\sigma_{\bi}},\qquad
 \sigma_{\bi}\frac{\del}{\del\sigma_{\bj}}\quad(\bi\leq\bj),\qquad
 \sigma_{M}\frac{\del}{\del\sigma_{M}},\\
\frac{\del^{2}}{\del\sigma_{\bk}\del\sigma_{\bl}},\quad
 \sigma_{i}\frac{\del^{2}}{\del\sigma_{\bk}\del\sigma_{\bl}}\quad
 (i\leq\bk+\bl),\quad
\sigma_{i}\sigma_{j}\frac{\del^{2}}{\del\sigma_{\bk}\del\sigma_{\bl}}\quad
 (i+j\leq\bk+\bl),\\
\sigma_{M}\frac{\del^{2}}{\del\sigma_{M}\del\sigma_{\bk}},\qquad
 \sigma_{M}\sigma_{\bi}\frac{\del^{2}}{\del\sigma_{M}\del\sigma_{\bk}}\quad
 (\bi\leq\bk),\\
\frac{\del}{\del\sigma_{M}}\biggl(\sigma_{M}\frac{\del}{\del\sigma_{M}}
 -\lambda\biggr),\qquad \sigma_{i}\frac{\del}{\del\sigma_{M}}
 \biggl(\sigma_{M}\frac{\del}{\del\sigma_{M}}-\lambda\biggr).
\end{gather}
\end{subequations}
The most general quasi-solvable operator of (at most) second-order
$\tcH^{(C,C')}$ is again obviously obtained by the linear combination of
all the above operators \eqref{eqs:opsCC'}. It should be remarked that
the operator $\tcH^{(C,C')}$ is always solvable since all the operators
in Eqs.~\eqref{eqs:opsCC'} preserve the following infinite flag of
finite dimensional linear spaces:
\begin{equation}
\label{eq:flagC}
\begin{array}{cccccl}
\tcV_{\cN_{1}\!,1;M}^{(C)} & \subset & \tcV_{\cN_{1}\!,2;M}^{(C)}
 & \subset\dots\subset & \tcV_{\cN_{1}\!,\cN_{2};M}^{(C)} & \subset\cdots\\
\cup & & \cup & & \cup & \\
\tcV_{\cN_{1}\!,1;M}^{(C')} & \subset & \tcV_{\cN_{1}\!,2;M}^{(C')}
 & \subset\dots\subset & \tcV_{\cN_{1}\!,\cN_{2};M}^{(C')} & \subset\cdots,
\end{array}
\end{equation}
for all $\cN_{1}=1,2,3,\dots$. In particular, the type C$'$ gauged
Hamiltonian \eqref{eq:gC'Ham} always belongs to type C, as has been mentioned
at the end of Section~\ref{sec:newtC}, thus always preserves the infinite
flag of the spaces \eqref{eq:flagC} for all $\cN_{1}=1,2,3,\dots$.
Table~\ref{tb:class} indicates that the rational, hyperbolic, trigonometric
$BC$ type, and hyperbolic, trigonometric $A$ type Calogero--Sutherland
systems have this intriguing property.

Furthermore, it is also interesting to note that the most general
operator of (at most) second-order $\tcH^{(A,A')}$ which preserves both
the type A and A$'$ spaces for a given $\cN=n$ always preserves the type
A$'$ spaces for all $\cN=1,2,3,\dots$ but does not the type A spaces
for any $\cN\neq n$ as far as the operator given by Eq.~\eqref{eq:2nd3CC'}
is included in $\tcH^{(A,A')}$. As a consequence, the operator
$\tcH^{(A,A')}$ is always solvable but the infinite flag of finite spaces
preserved by it has the different structure from Eq.~\eqref{eq:flagA}
as the following:
\begin{equation}
\begin{array}{cccl}
 & & \tcV_{n;M}^{(A)} & \\
 & & \cup & \\
\tcV_{1;M}^{(A')} & \subset\dots\subset
 \tcV_{n-1;M}^{(A')}\subset & \tcV_{n;M}^{(A')} & \subset
 \tcV_{n+1;M}^{(A')}\subset\cdots.
\end{array}
\end{equation}
Only when the operator \eqref{eq:2nd3CC'} does not exist in $\tcH^{(A,A')}$,
the infinite flag of finite spaces preserved by it has the same structure
as Eq.~\eqref{eq:flagA}.
The discussion we have made so far surely shows that linear spaces of
monomial type which can be preserved by quantum many-body systems have
far richer structure than those preserved by one-body systems.
The structure we have revealed in this article would be just the tip of
the iceberg.

Finally, we would like to recall the fact that the most general type C$'$
quasi-solvable operator \eqref{eq:tC'op2} preserves separately the subspaces
in the type C$'$ space \eqref{eq:mtypC'} due to the restriction
\eqref{eq:restr}. It indicates the existence of \emph{irregular} type C$'$
operators which do not preserve them separately when the restriction
\eqref{eq:restr} is omitted, as in the case of type C \cite{Ta05a}.
These issues on irregular type C$'$ together with irregular type C would
be reported elsewhere.

\begin{ack}
  This work was partially supported by a Spanish Ministry of
  Education, Culture and Sports research fellowship.
\end{ack}

\appendix

\section{Formulas}
\label{sec:formu}

In this appendix, we summarize some useful formulas on the elementary
symmetric polynomials.
\begin{align}
\label{eq:form1}
\sum_{k=0}^{M}(-1)^{k}z_{i}^{M-k}\sigma_{k}
 &=\sum_{k=0}^{M}(-z_{i})^{k}\sigma_{M-k}=0\quad(i=1,\dots,M),\\
\label{eq:form2}
\sum_{k=1}^{M}(-1)^{k+1}k z_{i}^{M-k-1}\sigma_{k}
 &=\prod_{j(\neq i)}^{M}(z_{i}-z_{j})\quad(i=1,\dots,M),\\
\label{eq:form3}
\frac{\del\sigma_{k}}{\del z_{i}}
 &=\sum_{l=1}^{k}(-z_{i})^{l-1}\sigma_{k-l}\quad(i,k=1,\dots,M),\\
\label{eq:chain}
\frac{\del}{\del\sigma_{k}}&=\sum_{i=1}^{M}
 \frac{(-1)^{k+1}z_{i}^{M-k}}{\prod_{j(\neq i)}^{M}(z_{i}-z_{j})}
 \frac{\del}{\del z_{i}}\quad (k=1,\dots,M).
\end{align}
The first formula \eqref{eq:form1} is readily derived from the following
identity:
\[
\prod_{i=1}^{M}(z-z_{i})=\sum_{k=0}^{M}(-1)^{k}z^{M-k}\sigma_{k}.
\]
The second one \eqref{eq:form2} is easily proved inductively. The third
one \eqref{eq:form3} is derived from the repeated application of
the following formula:
\[
\frac{\del\sigma_{k}}{\del z_{i}}=\sigma_{k-1}-z_{i}
 \frac{\del\sigma_{k-1}}{\del z_{i}}.
\]
The derivation of the fourth one \eqref{eq:chain} is as follows.
Taking the differential of Eq.~\eqref{eq:form1}, we have
\[
\dd z_{i}=\sum_{k=1}^{M}\frac{(-1)^{k+1}z_{i}^{M-k}}{\sum_{l=0}^{M-1}
 (-1)^{l}(M-l)z_{i}^{M-l-1}\sigma_{l}}\dd\sigma_{k}\quad(i=1,\dots,M).
\]
From Eq.~\eqref{eq:form1} the denominator in the r.h.s. of the above
equation reads,
\[
\sum_{l=0}^{M-1}(-1)^{l}(M-l)z_{i}^{M-l-1}\sigma_{l}
 =\sum_{l=1}^{M}(-1)^{l+1}lz_{i}^{M-l-1}\sigma_{l}.
\]
Hence, applying Eq.~\eqref{eq:form2} we obtain the formula \eqref{eq:chain}.

For the derivation of the formulas below, see Appendix~B in Ref.~\cite{Ta04}.
\begin{subequations}
\label{eqs:formBC}
\begin{align}
\sum_{i=1}^{M}\frac{\del}{\del z_{i}}&=\sum_{k=1}^{M}(M-k+1)\sigma_{k-1}
 \frac{\del}{\del\sigma_{k}},\\
\sum_{i=1}^{M}z_{i}\frac{\del}{\del z_{i}}&=\sum_{k=1}^{M}
 k\sigma_{k}\frac{\del}{\del\sigma_{k}},\\
2\sum_{i\neq j}^{M}\frac{1}{z_{i}-z_{j}}\frac{\del}{\del z_{i}}
 &=-\sum_{k=1}^{M}(M-k+1)(M-k+2)\sigma_{k-2}\frac{\del}{\del\sigma_{k}},\\
2\sum_{i\neq j}^{M}\frac{z_{i}}{z_{i}-z_{j}}\frac{\del}{\del z_{i}}
 &=-\sum_{k=1}^{M}(M-k)(M-k+1)\sigma_{k-1}\frac{\del}{\del\sigma_{k}},\\
2\sum_{i\neq j}^{M}\frac{z_{i}^{2}}{z_{i}-z_{j}}\frac{\del}{\del z_{i}}
 &=-\sum_{k=1}^{M}k(2M-k-1)\sigma_{k}\frac{\del}{\del\sigma_{k}}.
\end{align}
\end{subequations}

\bibliography{refsels}

\begin{thebibliography}{10}
\expandafter\ifx\csname url\endcsname\relax
  \def\url#1{{\tt #1}}\fi
\expandafter\ifx\csname urlprefix\endcsname\relax\def\urlprefix{URL }\fi
\providecommand{\eprint}[2][]{\url{#2}}

\bibitem{TU87}
A.~V. Turbiner and A.~G. Ushveridze, Phys. Lett. A 126 (1987) 181.

\bibitem{Us94}
A.~G. Ushveridze, {Q}uasi-exactly {S}olvable {M}odels in {Q}uantum {M}echanics
  (IOP Publishing, Bristol, 1994).

\bibitem{Tu88}
A.~V. Turbiner, Commun. Math. Phys. 118 (1988) 467.

\bibitem{RT95}
W.~R{\"u}hl and A.~Turbiner, Mod. Phys. Lett. A 10 (1995) 2213.
\newblock \eprint{hep-th/9506105}.

\bibitem{Ca71}
F.~Calogero, J. Math. Phys. 12 (1971) 419.

\bibitem{Su71}
B.~Sutherland, Phys. Rev. A 4 (1971) 2019.

\bibitem{MRT96}
A.~Minzoni, M.~Rosenbaum, and A.~Turbiner, Mod. Phys. Lett. A 11 (1996) 1977.
\newblock \eprint{hep-th/9606092}.

\bibitem{BTW98}
L.~Brink, A.~Turbiner, and N.~Wyllard, J. Math. Phys. 39 (1998) 1285.
\newblock \eprint{hep-th/9705219}.

\bibitem{HS99}
X.~Hou and M.~Shifman, Int. J. Mod. Phys. A 14 (1999) 2993.
\newblock \eprint{hep-th/9812157}.

\bibitem{GGR00}
D.~G{\'o}mez-Ullate, A.~Gonz{\'a}lez-L{\'o}pez, and M.~A. Rodr{\'\i}guez, J.
  Phys. A: Math. Gen. 33 (2000) 7305.
\newblock \eprint{nlin.SI/0003005}.

\bibitem{Ta04}
T.~Tanaka, Ann. Phys. 309 (2004) 239.
\newblock \eprint{hep-th/0306174}.

\bibitem{In83}
V.~I. Inozemtsev, Phys. Lett. A 98 (1983) 316.

\bibitem{IM85}
V.~I. Inozemtsev and D.~V. Meshcheryakov, Lett. Math. Phys. 9 (1985) 13.

\bibitem{In89}
V.~I. Inozemtsev, Lett. Math. Phys. 17 (1989) 11.

\bibitem{PT95}
G.~Post and A.~Turbiner, Russ. J. Math. Phys. 3 (1995) 113.
\newblock \eprint{funct-an/9307001}.

\bibitem{GT04}
A.~Gonz{\'a}lez-L{\'o}pez and T.~Tanaka, Phys. Lett. B 586 (2004) 117.
\newblock \eprint{hep-th/0307094}.

\bibitem{GT05}
A.~Gonz{\'a}lez-L{\'o}pez and T.~Tanaka, J. Phys. A: Math. Gen. 38 (2005) 5133.
\newblock \eprint{hep-th/0405079}.

\bibitem{Ta05a}
T.~Tanaka, Ann. Phys. 316 (2005) 187.
\newblock \eprint{hep-th/0407275}.

\bibitem{Tu94}
A.~V. Turbiner, Contemp. Math. 160 (1994) 263.

\bibitem{BB98}
C.~M. Bender and S.Boettcher, J. Phys. A: Math. Gen. 31 (1998) L273.
\newblock \eprint{physics/9801007}.

\bibitem{BM05}
C.~M. Bender and M.~Monou, 2005.
\newblock New quasi-exactly solvable sextic polynomial potentials.
\newblock Preprint, \eprint{quant-ph/0501053}.

\bibitem{CRT98}
A.~Capella, M.~Rosenbaum, and A.~Turbiner, Int. J. Mod. Phys. A 13 (1998) 3885.
\newblock \eprint{solv-int/9707005}.

\bibitem{Ta03c}
T.~Tanaka, 2003.
\newblock {$\mathcal N$}-fold supersymmetry and quasi-solvability.
\newblock To appear in Progress in Mathematical Physics Research (Nova Science
  Publishers, New York).

\bibitem{OP83}
M.~A. Olshanetsky and A.~M. Perelomov, Phys. Rep. 94 (1983) 313.

\bibitem{KM98}
A.~Khare and B.~P. Mandal, J. Math. Phys. 39 (1998) 5789.
\newblock \eprint{quant-ph/9803003}.

\end{thebibliography}
\bibliographystyle{npb}



\end{document}